\documentclass[useAMS,usenatbib]{mn2e}

\usepackage{newtxtext,newtxmath}

\usepackage[T1]{fontenc}
\usepackage{ae,aecompl}

\usepackage{graphicx}	
\usepackage{amsmath}	
\usepackage{amssymb}	
\usepackage{morefloats}
\usepackage{subfig}
\usepackage{rotating}
\usepackage{pdflscape}

\usepackage{amsfonts}
\usepackage{pifont}

\usepackage{color} 
\definecolor{purple}{RGB}{160,32,240}
\usepackage {comment}
\usepackage{url}

\newcommand{\rockstar}{\textsc{Rockstar}}
\newcommand{\ctrees}{\textsc{Consistent Trees}}
\newcommand{\vmax}{v_\mathrm{max}}

\def\aap{\mbox{A\&A}}
\def\apjl{\mbox{ApJ}}
\def\apjs{\mbox{ApJS}}

\graphicspath{{./final_plots/}}

\def\msun{\mbox{M$_{\odot}$}}

\def\mvir{\mbox{$M_{\rm vir}$}}

\def\cvir{\mbox{$c_{\rm vir}$}}

\def\mpeak{\mbox{$M_{\rm peak}$}}

\def\vmax{\mbox{$V_{\rm max}$}}

\defcitealias{RDA12}{RDA12}

\newcommand{\doff}{D_{\rm off}}
\newcommand{\lambdap}{\lambda_{\rm B}}
\newcommand{\lambdaB}{\lambda_{\rm B}}
\newcommand{\lambdaP}{\lambda_{\rm P}}

\newcommand{\cnfw}{C_{\mathrm{NFW}}}

\newcommand{\rvir}{R_{\mathrm{vir}}}
\newcommand{\rs}{R_{s}}
\newcommand{\rfive}{R_{500c}}
\newcommand{\pvir}{P_{\rvir}}
\newcommand{\pfive}{P_{\rfive}}

\newcommand{\almm}{a_{\mathrm{LMM}}}

\newcommand{\xoff}{X_{\mathrm{off}}}
\newcommand{\tf}{\mathrm{TF}}
\newcommand{\tu}{T/|U|}
\newcommand{\mar}{\dot{M}/M}
\def\ltsima{$\; \buildrel < \over \sim \;$}    
\def\lesssim{\lower.5ex\hbox{\ltsima}}           
\def\gtsima{$\; \buildrel > \over \sim \;$}    
\def\grtsim{\lower.5ex\hbox{\gtsima}}           

\title[Tidal Stripping and Post-Merger Relaxation]{Tidal Stripping and Post-Merger Relaxation of Dark Matter Halos:  Causes and Consequences of Mass Loss}

\author[]{Christoph T. Lee,$^{1}$\thanks{E-mail: chtlee@ucsc.edu}, Joel R. Primack$^1$, Peter Behroozi$^2$, Aldo Rodr\'iguez-Puebla$^3$, 
\newauthor Doug Hellinger$^1$, 
Avishai Dekel$^4$\\
$^{1}$ Physics Department, University of California, Santa Cruz, CA 95064, USA\\
$^2$ Department of Astronomy, University of Arizona, 933 N Cherry Ave, Tucson, AZ 85719 USA \\
$^3$ Instituto de Astronom\'ia, Universidad Nacional Aut\'onoma de M\'exico, A. P. 70-264, 04510, M\'exico, D.F., M\'exico \\
$^4$ Center for Astrophysics and Planetary Science, Racah Institute of Physics, The Hebrew University, Jerusalem 91904, Israel \\
}

\date{Accepted XXX. Received YYY; in original form ZZZ}

\pubyear{2017}

\begin{document}

\label{firstpage}
\pagerange{\pageref{firstpage}--\pageref{lastpage}}

\maketitle

\begin{abstract}
We study the properties of distinct dark matter halos (i.e., those that are not subhalos) that have a final virial mass $\mvir$ at $z = 0$ less than their peak mass ($\mpeak$) in the Bolshoi-Planck cosmological simulation. We identify two primary causes of halo mass loss: relaxation after a major merger and tidal stripping by a massive neighbouring halo.  Major mergers initially boost $\mvir$ and typically cause the final halo to become more prolate and less relaxed and to have higher spin and lower NFW concentration.  As the halo relaxes, high energy material from the recent merger gradually escapes beyond the virial radius, temporarily resulting in a net negative accretion rate that reduces the halo mass by $5-15\%$ on average.  Halos that experience a major merger around $z = 0.4$ typically reach a minimum mass near $z = 0$.  Tidal stripping mainly occurs in dense regions, and it causes halos to become less prolate and have lower spins and higher NFW concentrations.  Tidally stripped halos often lose a large fraction of their peak mass ($> 20\%$) and most never recover (or even reattain a positive accretion rate).  Low mass halos can be strongly affected by both post-merger mass loss and tidal stripping, while high mass halos are predominantly influenced by post-merger mass loss and show few signs of significant tidal stripping.  
\end{abstract}




\section{Introduction}

In the modern $\Lambda$CDM standard cosmological paradigm, galaxies form and evolve within dark matter halos \citep[reviewed in][]{FrenkWhite12,Primack12}.  Determining the evolving properties of the halos is therefore important for understanding the formation of galaxies in the expanding universe.  Ever since simulations were first able to resolve dark matter halos \citep[e.g.,][]{NFW96} there have been studies of the abundance and other properties of the halos---most recently, using the Planck \citep{Planck13,Planck15} cosmological parameters \citep[][and references therein]{Klypin16,HaloDemographics}.  The general expectation has been that dark matter halos typically grow in mass with cosmic time, for example with the mass at any redshift $z$ approximately given by $M(z) = M(0) \exp(-\alpha z)$, where $\alpha \sim 0.8$ is fit to individual halo growth trajectories \citep[e.g.][and references therein]{Wechsler02,Dekel13,HaloDemographics,RP17}.  
However, \citet{Lee17} showed that in regions of high dark matter density, most low-mass halos are actually losing mass due to tidal stripping.
But although some halos lose rather than gain mass, until the present paper there has not been any detailed investigation of the causes and consequences of halo mass loss.  As we will show, there are two main causes of halo mass loss: relaxation after major mergers and tidal stripping.  

After a halo major merger (i.e., with the merging halos having mass ratio greater than 0.3 to 1), the virial ratio of the kinetic and potential energies $T/|U|$ of the resulting halo is often initially considerably larger than the virial value 0.5.  As the halo relaxes toward virial equilibrium, a small fraction of its mass, typically $\sim$10\%, moves beyond the virial radius.  This includes a large fraction of the higher angular momentum material, and as a consequence the halo spin parameter $\lambda$ decreases, as shown by \citet{D'OnghiaNavarro07}.
An example of halo mass loss after a major merger was demonstrated and briefly discussed in \citet[][Section 5.2]{Notts}, and the phenomenon was also mentioned in \citet{BehrooziLoebWechsler}.  As we will see, this post-merger mass loss phenomenon is rather common.

Tidal stripping of subhalos has been extensively discussed in the literature \citep[e.g.][and refererences therein]{vandenBosch17}, but here we concentrate on distinct halos.  Tidal stripping of distinct dark matter halos that are close to (or pass through) more massive halos has been discussed in the literature \citet{Hahn09,BehrooziLoebWechsler,Behroozi14,Hearin16}.  It can lead to more than 10\% mass loss, sometimes much more.  This affects only a relatively small fraction of the halos except in dense regions, where most halos with mass $\lesssim 10^{12} M_\odot$ suffer significant mass loss.  

Interestingly, the effects of these two halo mass loss mechanisms are nearly orthogonal.  As we will show, halos suffering mass loss after a major merger typically have lower concentration and higher spin and prolateness than average, while halos that are significantly tidally stripped have higher concentration and lower spin and prolateness.  

This paper is organized as follows.  In \S2 we discuss the simulations and methods used, including definitions of key concepts that we use, including tidal force TF, halo concentration $C_{\rm NFW}$, and prolateness $P$.  \S3 discusses the frequency, causes, and consequences of mass loss, with \S3.1 emphasizing halos mass loss after major mergers and \S3.2 focussing on tidal stripping.  We recommend that readers skip ahead to Figures 13 and 14, which show typical behaviours of many halo properties after major mergers and after tidal stripping.  
\S4 presents our Discussion and Conclusions.

\section{Simulations and Method}
\label{sec:Sims}

In this paper we study the halos in the Bolshoi-Planck simulation \citep{Klypin16,HaloDemographics}, which have been analyzed using the \rockstar\ halo finder \citep{ROCKSTAR} and the \ctrees\ formalism \citep{ConsTrees} for constructing halo merger trees.  
The cosmological parameters for the Bolshoi-Planck simulation were $\Omega_{\Lambda,0}=0.693,\Omega_{\rm M,0}=0.307,\Omega_{\rm B,0}=0.048, h=0.678, n_s=0.96$  and $\sigma_8=0.823$.
The halo masses studied are given by 
\begin{equation}
\mvir = \frac{4\pi}{3} \Delta_{\rm vir} \rho_{\rm m} R_{\rm vir}^3 ,
\end{equation}
where $\Delta_{\rm vir}$ is given by the \citet{BryanNorman} fitting formula
$\Delta_{\rm vir}(z) = (18\pi^2 + 82x - 39x^2)/\Omega(z)$, where $\Omega(z) $ is the ratio of mean matter density $\rho_{\rm m}$ to critical density $\rho_{\rm c}$ at redshift $z$, and $x \equiv \Omega(z) -1$. 
{\color{black}Using other halo mass definitions \citep[e.g., splashback radius][]{Splashback15} 
and/or other halo finders would lead to different results \citep[e.g.][]{Notts}, but we do not consider such alternatives in this paper since doing so would require reanalysis of the many timesteps of a large simulation.}

The outputs from the 178 saved timesteps of the Bolshoi-Planck simulation analyzed by \rockstar\ are discussed in \citet[][especially the Appendices]{HaloDemographics}, and they can be downloaded from the UCSC Hyades astronomical computer system.\footnote{\url{http://hipacc.ucsc.edu/Bolshoi/MergerTrees.html}} 
Here we summarize the definitions of the halo properties that we study in this paper.

\begin{itemize}

\item{NFW concentration} $\cnfw$ is defined as 
\begin{equation}
\cnfw = \frac{R_{\rm vir}}{R_s} ,
\end{equation}
where the  \citet[][NFW]{NFW96} profile is given by
\begin{equation}
\rho_{\rm NFW}(r) = \frac{4\rho_s}{(r/R_s)(1+r/R_s)^2} .
\end{equation}
The scale radius $R_s$ is the radius where the logarithmic slope of the density profile is -2. The NFW profile is completely characterized by two parameters, for example $\rho_s$ and $R_s$, or alternatively the halo mass $\mvir$ and concentration parameter $\cvir$. 

\item{Maximum circular velocity} $\vmax$ is the maximum of $[G M(<r)/r]^{1/2}$ at any radius $r<\rvir$, where $M(<r)$ is the mass enclosed by spherical radius $r$.

\item{Bullock spin parameter} $\lambdaB$ is defined following \citet{Bullock+2001} as
\begin{equation}
\lambdaB = \frac{J}{\sqrt{2} M_{\rm vir} V_{\rm vir} R_{\rm vir}},
\label{lambdaB}
\end{equation}
The distribution and redshift dependence of $\lambdaB$ is discussed in \citet{HaloDemographics} and \citet[][Appendix B]{Somerville17} and compared with that of the \citet{Peebles69} spin parameter
	\begin{equation}
		\lambda_{\rm P} = \frac{J  | E |^{1/2}} { G M_{\rm vir}^{5/2}},
	\label{lambdaP}
	\end{equation}
where $J$ and $E$ are the total angular momentum and the total energy of a halo of mass \mvir. 

\item{Scale factor of the last major merger} $\almm$ is the scale factor $a=(1+z)^{-1}$ of the most recent merger with halo mass ratio greater than 0.3 to 1 along the halo's largest progenitor track.

\item{Tidal Force (TF)} experienced by a halo is calculated as the strongest tidal force from any nearby halo, in dimensionless units ($R_\mathrm{halo} / R_\mathrm{Hill}$).   The Hill radius \citep[see, e.g.,][]{Hahn09,Hearin16} is the radius within which material can remain gravitationally bound to a secondary halo.  It is given by $R_\mathrm{Hill} = D(M_\mathrm{sec}/3M_\mathrm{prim})^{1/3} = R_\mathrm{sec}(D/3^{1/3} R_\mathrm{prim})$, where the primary (secondary) halo is the larger-virial-radius (smaller-virial-radius) halo of a pair whose centers are separated by distance $D$.  

\item{Prolateness} is defined as 
\begin{equation}
P \equiv 1-\frac{1}{\sqrt[]{2}}\left[\left(\frac{b}{a}\right)^{2} + \left(\frac{c}{a}\right)^{2}\right]^{1/2},
\label{eq:prolateness}
\end{equation}
such that $1-P$ is the magnitude of the vector $(\frac{b}{a},\frac{c}{a})$ normalized by $\frac{1}{\sqrt{2}}$, where $a$, $b$, and $c$ are the lengths of the largest, second largest, and smallest triaxial ellipsoid axes, respectively.  Prolateness ranges from 0 (perfect sphere) to 1 (maximally elongated, i.e. a needle), with most halos falling somewhere in the range of $0.2 - 0.6$.  We will plot the Prolateness of halos both within the virial radius $\rvir$ and the radius $R_{500c}$ enclosing average density 500 times critical density $\rho_c = 3H^2/(8\pi G)$, where $H=\dot a/a$ is the Hubble parameter.

\item{} $\xoff$ is the offset of the density peak of a halo from its center of mass within $\rvir$, in units of $h^{-1}$ kpc. 

\item{} $\doff = \xoff/\rvir$.

\item{Virial ratio}  $T/|U|$ is the ratio of kinetic energy $T$ to potential energy $|U|$.  According to the virial theorem, a fully relaxed halo has $T/|U|=0.5$.

\item{Environmental density} $\rho_{\sigma}$ is the local environment density smoothed on different scales ($\sigma$) and $\rho_{\mathrm{avg}}$ is the average density of the simulation.  (See \citep{Lee17} for details.)

\end{itemize}

As in \citet{Lee17}, in this paper we will plot halos in four mass bins, corresponding to $\log_{10} \mvir / (h^{-1} M_\odot) = 11.2 \pm 0.375$, $11.95  \pm 0.375$, $12.7  \pm 0.375$, and $13.45 \pm 0.375$.  {\color{black}The lowest mass $z = 0$ halos in our analysis still contain roughly $\sim 450$ particles to minimize noise introduced from low particle counts.}  As shown in \citet{HaloDemographics}, $\mvir = 10^{12.7} h^{-1}M_\odot$ is the mass of $1\sigma$ fluctuations just collapsing at $z=0$, and the other mass bins are separated by $10^{0.75} h^{-1}M_\odot$.  As in \citet{Lee17}, we quote the environmental density $\rho_{\sigma}$ smoothed at 1 $h^{-1}$Mpc for halos in the lowest mass bin, 2 $h^{-1}$Mpc in the next mass bin, 4 $h^{-1}$Mpc  in the next mass bin, and 8 $h^{-1}$Mpc in the highest mass bin; these radii are large enough that the central halo has little effect on the density.

\section{How mass loss affects halo properties}
\label{sec:Mass Loss}

\begin{figure}
	\centering
	\includegraphics[trim=0 3 5 10, clip, width=\columnwidth]{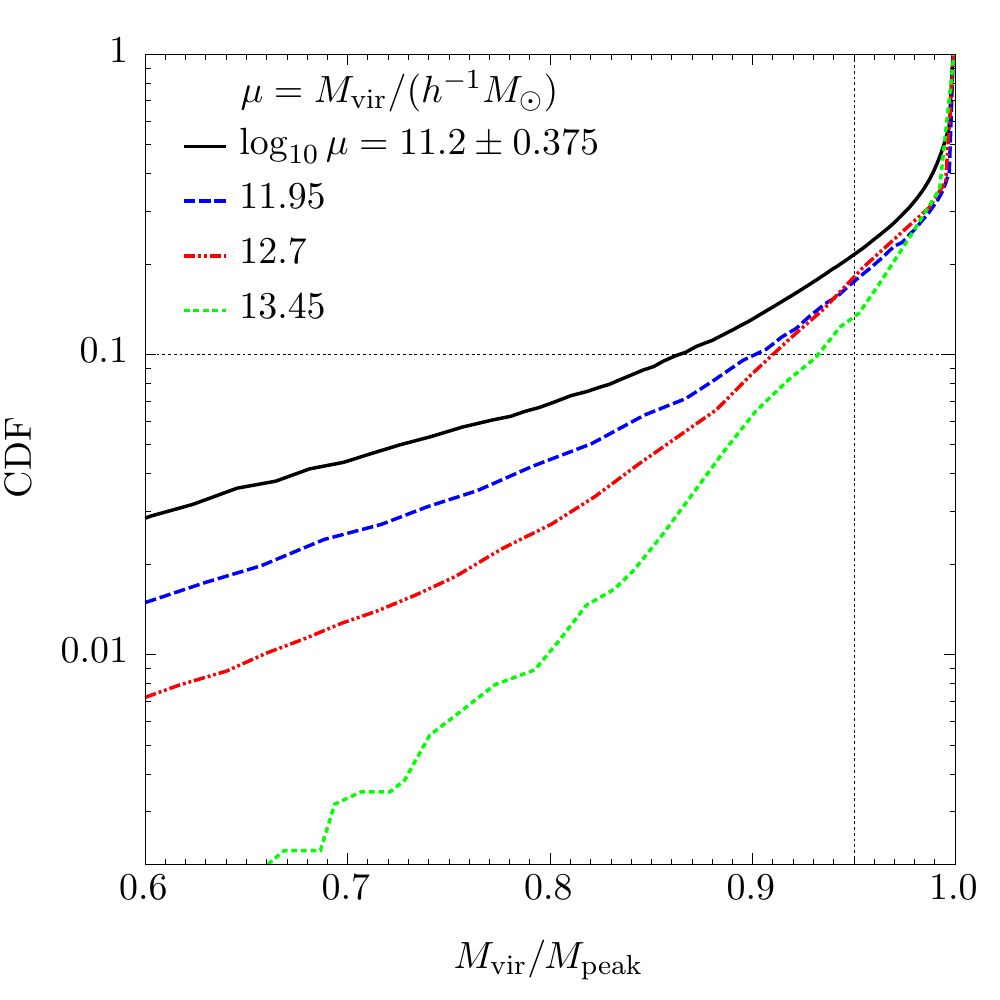}
    \caption{Cumulative distribution function of mass loss fraction for distinct halos. {\color{black}We divide our halos into 4 mass bins of width $\Delta \log_{10} \mu = 0.75$.} Lower mass halos have typically lost more mass than higher mass halos.  The fraction of halos that have experienced appreciable mass loss (greater than $5\%$ since their peak mass) ranges from roughly $12\%$ (highest mass bin) to $22\%$ (lowest mass bin). Roughly $5\%$ of low mass halos have experienced dramatic mass loss ($ > 30\%$ since $\mpeak$), while very few high mass halos have.}
    \label{fig:stripping_cdf}
\end{figure}

\begin{figure*}
	\centering
	\includegraphics[trim=60 50 110 30, clip, width=0.95\textwidth]{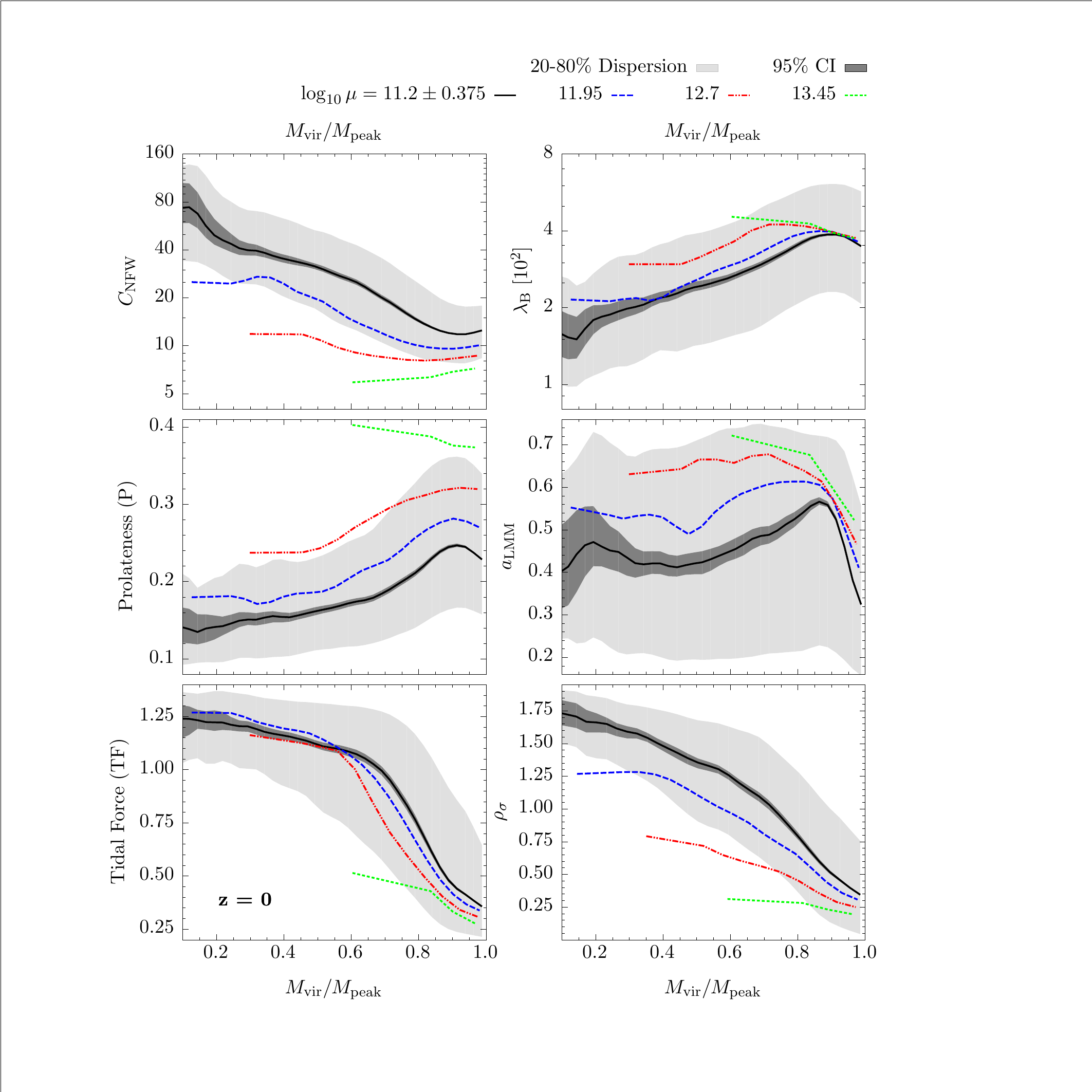}
    \caption{Median $z = 0$ relations between mass loss fraction and NFW concentration ($\cnfw$), Bullock spin parameter ($\lambdap$), prolateness (P), scale factor of last major merger ($\almm$), tidal force (TF), and local density ($\rho_{\sigma}$).  {\color{black}We use the same mass bin definitions as in Fig. \ref{fig:stripping_cdf}.} We include only distinct halos.  Light grey shading reflects the 20-80th percentile dispersion, while dark grey shading indicates the $95\%$ confidence interval on the median, shown only for the lowest mass bin.  Halos that have lost a small amount of mass ($5-15\%$ for low mass halos) tend to have lower concentrations, higher spin parameters, and are more prolate compared to halos that have not lost mass.  Conversely, halos that have experienced more dramatic mass loss ($> 20\%$ for low mass halos) display the opposite behaviour: they have lower concentrations, lower spin parameters, and are more spherical compared to halos that have not lost mass.  Additionally, halos that have lost a small amount of mass are much more likely to have experienced a recent major merger, while halos that have experienced dramatic mass loss tend to experience higher tidal forces and live in significantly higher density regions than halos that have not lost mass.  These trends are most pronounced for lower mass halos.  The competing trends between weak and dramatic mass loss are subdued and the transition occurs at higher mass loss fractions for higher mass halos.}
    \label{fig:stripping_properties_correlation}
\end{figure*}

We first motivate our analysis of mass loss mechanisms and how they affect halo properties by looking at how common mass loss is among distinct halos at $z = 0$.  In Fig. \ref{fig:stripping_cdf}, we plot the cumulative distribution of the mass loss fraction ($\mvir/\mpeak$), which describes how much of a halos' maximum mass remains at $z = 0$.  For many halos this fraction is 1 (or, their maximum mass is their current mass).  We will consider halos to be ``diminished'' when they have experienced mass loss.  Halos that have lost more than $5\%$ of their mass since their peak mass are appreciably diminished.  We will generally consider halos that have lost less than $5\%$ of their mass since their peak mass to belong to the group of undiminished -- or normal -- halos.  This distinction between diminshed and normal halos is provided by the vertical dotted black line in Fig. \ref{fig:stripping_cdf}.  We see that the lowest mass bin is composed of roughly $22\%$ diminished halos, while the highest mass bin contains about $12\%$.  We haven't yet established at what degree of mass loss other halo properties are noticeably affected, but clearly a significant fraction of halos are potentially subject to these effects.  An additional point to establish is that mass loss is more common for lower mass halos.  High mass halos rarely experience dramatic mass loss, while roughly $5\%$ of low mass halos have lost more than $30\%$ of their mass since their peak mass.

To begin understanding how mass loss affects other halo properties, we plot in Fig \ref{fig:stripping_properties_correlation} relations between the mass loss fraction ($\mvir/\mpeak$) and several halo properties. Each panel presents medians of the given relation for each of the four halo mass bins (we use the same four mass bins throughout this analysis). For the lowest mass halos we show the $20-80\%$ dispersion with light grey shading and the $95\%$ confidence interval on the median in dark grey shading. Higher mass halos may have similar dispersions, but will have wider confidence intervals (not shown) due to having fewer halos. High mass halos also do not extend to very low values of $\mvir/\mpeak$ since they rarely lose more than about $30\%$ of their peak mass.  We observe similar trends in spin parameter, prolateness, and (inverted) concentration as a function of mass loss fraction. In each of these properties we identify a change in one direction coincident with a moderate amount of mass loss, along with a change in the opposite direction for heavily diminished halos. For halos in the lowest mass bin this turnaround occurs at a mass loss level of roughly $10\% \ \mpeak$; for higher mass halos the turnaround is more gradual and shifted towards higher levels of mass loss. Overall, we see that low mass halos that have lost between $5-15\% \ \mpeak$ have lower concentrations, higher spin parameters, and are more prolate than normal halos, while those that have lost more than $15\% \ \mpeak$ have increasingly higher concentrations, lower spin parameters, and are more spherical than normal halos.

Looking at the remaining panels of Fig. \ref{fig:stripping_properties_correlation}, we see that moderately diminished halos are far more likely to have had a recent major merger than normal halos.  At least for low mass halos, there is a strong peak in the typical $\almm$ coincident with a mass loss of about $15\% \ \mpeak$, while normal halos experienced very few recent major mergers.  This suggests that merging history may play an important role in influencing these halo properties, especially in the moderate mass loss regime, where we observe the strongest correlation with $\almm$.

Finally, in the bottom two panels of Fig. \ref{fig:stripping_properties_correlation}, we show the relations between tidal force, local density and mass loss fraction.  We see a monotonic relationship with mass loss fraction in both cases and for all mass bins. Halos that have lost more mass tend to be experiencing stronger tidal forces and are found in higher density regions than those that have lost less mass.  We note that the tidal force we've used is an approximation of the true tidal force, since for each halo we consider only the force from the single most tidally influential nearby halo.  The local density, however, is computed by smoothing the particle data from the simulation, providing an accurate indication of the density of the surrounding environment.  We expect that the plateauing of the tidal force for halos that have lost more that about $40\% \ \mpeak$ is an artifact of our approximation method, and that the true tidal force would scale more linearly with mass loss fraction, as does the local density.

Given these relationships between the intensity of halo mass loss and other halo properties, we found it natural to propose two distinct mass loss mechanisms:
\begin{enumerate}
	\item mass loss due to relaxation after a recent major merger, and
	\item mass loss due to tidal stripping in very high density environments.
\end{enumerate} These two mechanisms are not exclusive, and in (somewhat rare) cases can both be contributing to mass loss simultaneously.  To quantify these processes, we divide all halos in each mass bin into one of four groups, labelled as follows:
\begin{itemize}
	\item Tidal Stripping (TS): potentially subject to tidal stripping (has experienced tidal force $> 1$ since $\mpeak$), but has \textbf{not} experienced mass loss from a recent major merger ($\almm < 0.45$);
	\item Relaxation (R): \textbf{not} subject to tidal stripping, but potentially subject to mass loss during the relaxation period following a recent major merger ($\almm > 0.45$);
	\item Tidal Stripping + Relaxation (TS+R): potentially subject to \textbf{both} tidal stripping and relaxation mass loss following a recent major merger;
	\item Neither (N): does not satisfy conditions for either mass loss mechanism. 

\end{itemize}

{\color{black}Note that we have not yet imposed any selections based on mass loss directly.  Each of these groups contain halos that cover a broad range of mass loss levels, including zero mass loss.  In certain subsequent figures, we'll specifically select diminished halos (those that have lost $> 5\%$ of their peak mass) from these groups to emphasize the effects of each mass loss mechanism.}

\begin{figure}
	\centering
	\includegraphics[trim=70 70 35 35, clip, width=\columnwidth]{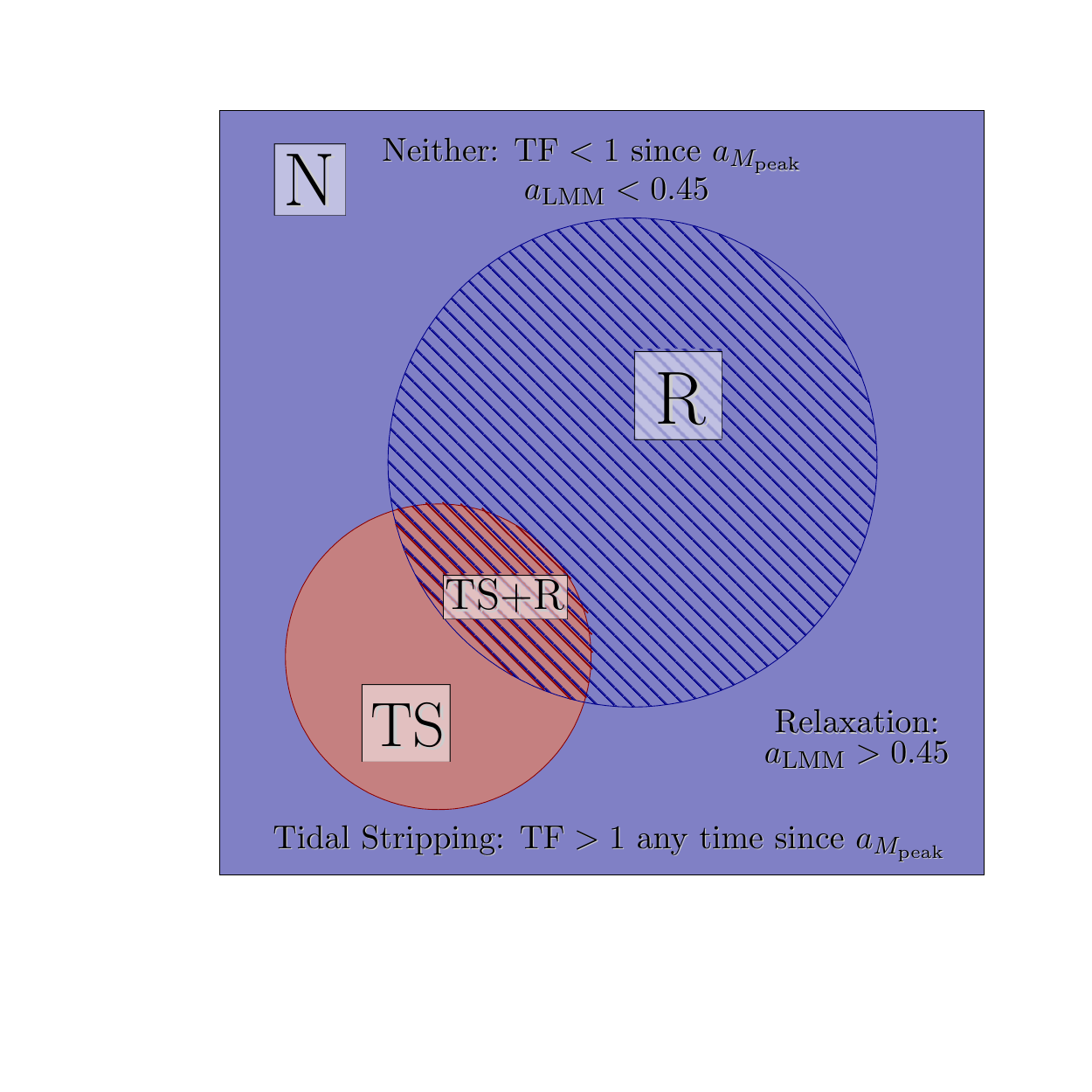}
    \caption{We present a visual representation of the four mass loss group labels we use in this analysis.  All halos in each mass bin are assigned to one of these groups according to their tidal force and major merger history.  TS (tidal stripping) halos are subject to tidal stripping only, R (relaxation) halos are subject to post major merger mass loss only, TS+R halos are subject to both tidal stripping and post-merger mass loss, and N (neither) halos do not satisfy the conditions for either mass loss mechanism.}
    \label{fig:mechanism_freq_venn}
\end{figure}

\begin{figure}
	\centering
	\includegraphics[trim=32 80 35 20, clip, width=\columnwidth]{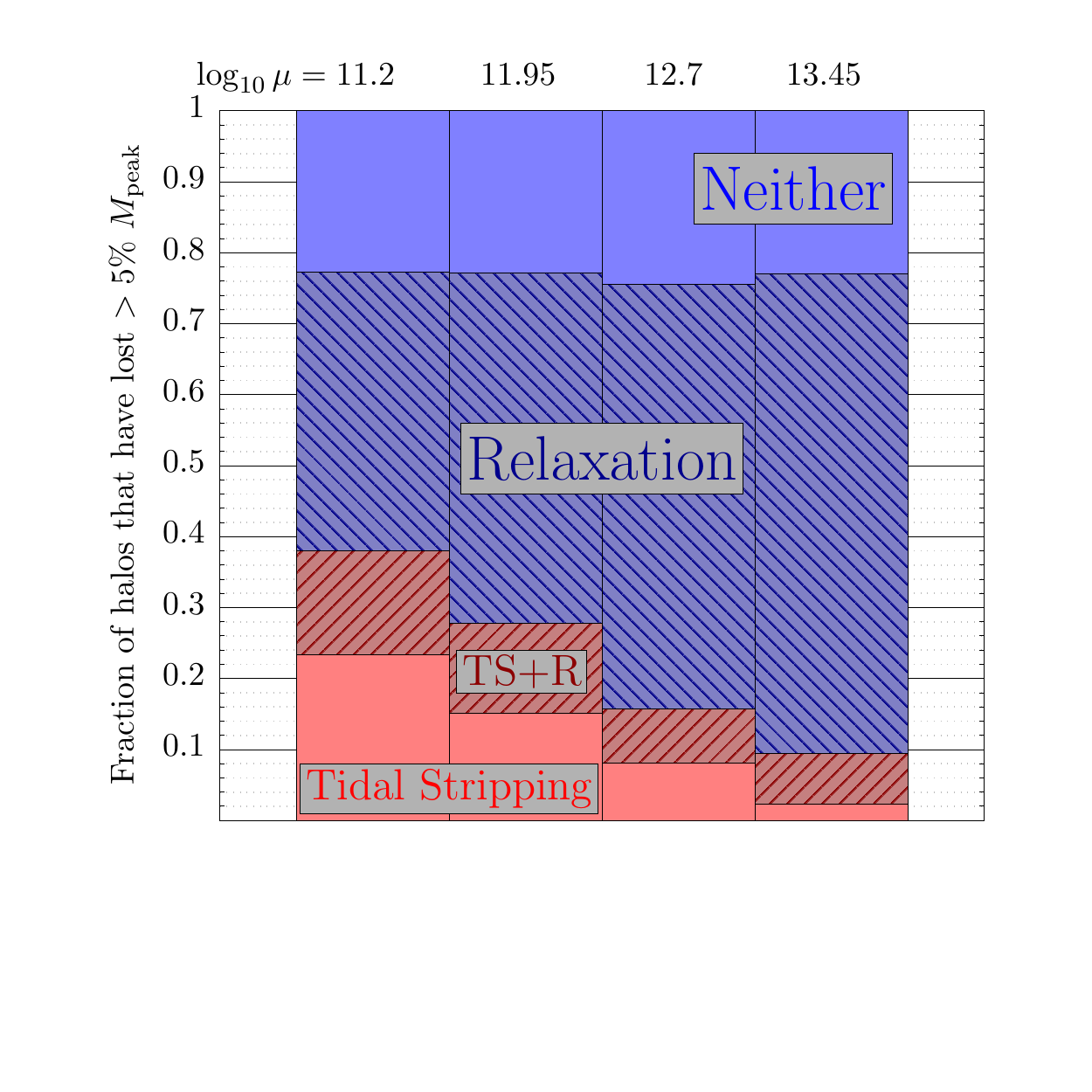}
    \caption{Fraction of halos in each mass loss group for different halo mass bins.  From each group, we show only the subset of distinct halos that have lost more than $5\%$ of their peak mass.  {\color{black}We use the same mass bin definitions as in Fig. \ref{fig:stripping_cdf}.} Group N halos are not apparently subject to the mass loss mechanisms we identify in this analysis, Group TS halos are potentially subject to tidal stripping but not major merger induced mass loss, Group R halos are potentially subject to mass loss following a major merger but not tidal stripping, and Group TS+R halos are potentially subject to both tidal stripping and major merger induced mass loss.  Diminished lower mass halos are much more likely to be found in Group TS (about $23\%$ of diminished low mass halos compared to only about $2\%$ of diminished high mass halos), but less likely to be in Group R (about $39\%$ of diminished low mass halos compared to about $67\%$ of diminished high mass halos). Altogether, about $77\%$ of diminished halos fall into one of the three mass loss groups (TS, R, and TS+R); the remaining $23\%$ of halos (group N) have likely lost mass via minor mergers.}
    \label{fig:mechanism_freq}
\end{figure}

In Fig. \ref{fig:mechanism_freq_venn}, we show a visual representation of how these groups are organized (with shape areas representative of the mass bin centred on $\log_{10} \mvir/\msun = 11.2$); additionally, in Fig. \ref{fig:mechanism_freq} we show how many diminished halos fall into each of these groups for each of the four mass bins in our analysis.  We find that low mass halos are the most likely to be potentially subject to tidal stripping alone (about $23\%$ of halos in our lowest mass bin are in group TS compared to only about $2\%$ of halos in the highest mass bin).  However, high mass halos are more commonly subject to major merger induced mass loss (about $67\%$ of halos in our highest mass bin are in group R compared to only about $39\%$ of low mass halos).  Few halos are potentially subject to both mass loss mechanisms (group TS+R), with the total ranging from $15\%$ of low mass halos to only $7\%$ of high mass halos. The combined fraction of halos that are potentially subject to these two mass loss mechanisms (groups R, TS, and TS+R) is about $77\%$ for all mass bins.  

%
%
%

\begin{figure*}
	\centering
	\includegraphics[trim=60 50 110 50, clip, width=.8\textwidth]{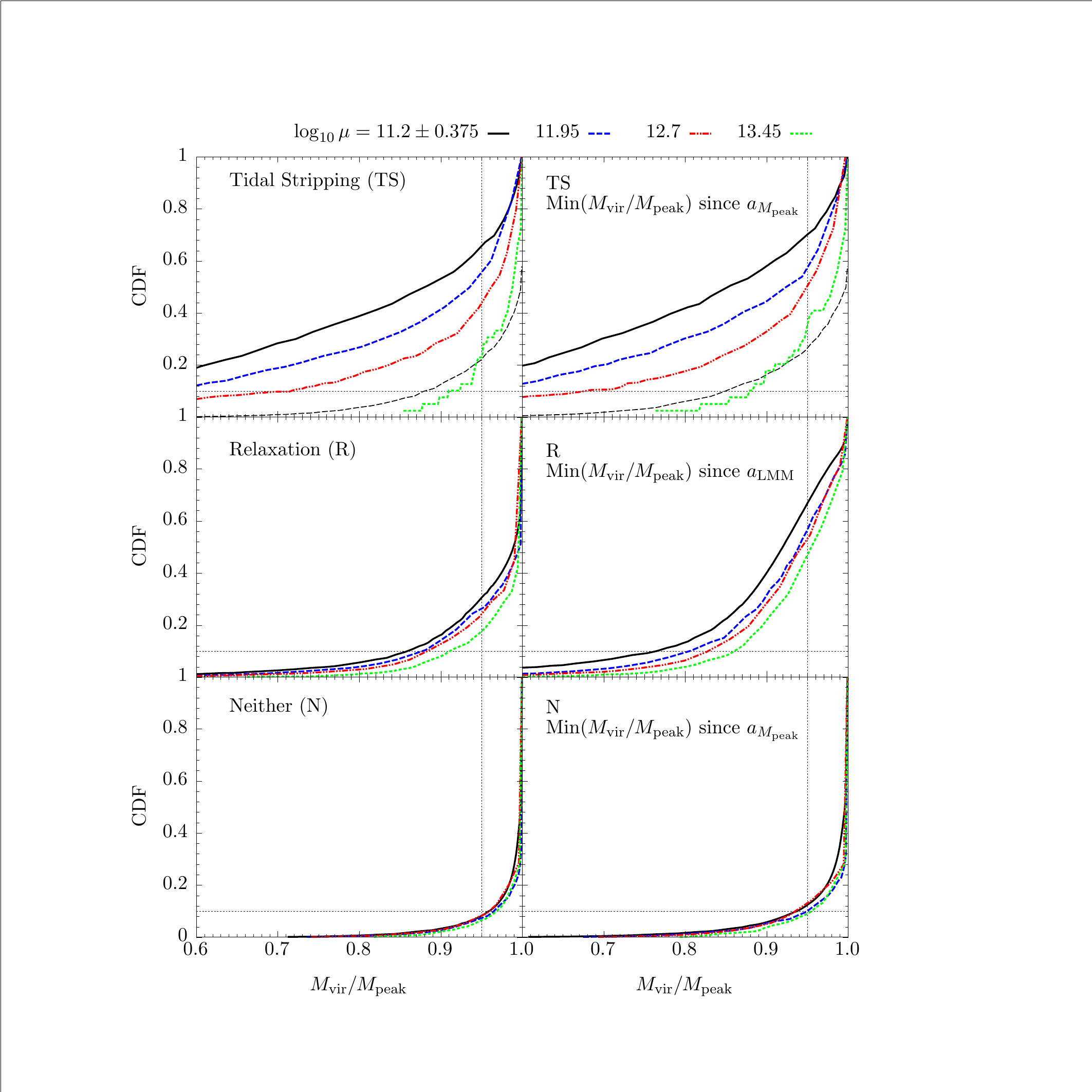}
    \caption{Cumulative distribution functions of mass loss fraction for distinct halos that are tidally stripped (group TS, top row), have had a recent major merger (group R, middle row), and are neither tidally stripped nor had a recent major merger (group N, bottom row). {\color{black}We use the same mass bin definitions as in Fig. \ref{fig:stripping_cdf}.} In the left column we define the mass loss fraction for each halo as the ratio of final halo mass at $z = 0$ to peak halo mass, while in the right column we instead compute the fraction using the minimum virial mass since the halo reached $\mpeak$ (or since $\almm$ for group R halos).  This allows us to compare the distribution of mass loss fractions at $z = 0$ compared to the distribution of peak mass loss for each group.  Different coloured lines represent different mass bins.  We see that TS halos experienced the greatest mass loss, with very little cumulative increase in peak mass loss (right column) compared to mass loss at $z = 0$ -- an indication that nearly all TS halos are actively losing mass at $z = 0$ and are not likely to recover.  Roughly $40\%$ of TS halos in the lowest mass bin experience a peak mass loss of greater than $20\%$.  We also see a strong mass dependence with TS halos; high mass halos experience far less tidal stripping than low mass halos.  For TS halos in the lowest mass bin only, we use the thin dashed black line to additionally show the distribution for those that have never been subhalos; to facilitate comparison this curve is not independently normalized, but rather uses the same normalization as the full mass bin (solid black curve).  We see that only about $35\%$ of diminished low mass halos have never been subhalos, and that almost all halos stripped of more than $20\%$ of their peak mass have been subhalos previously.  Only about $30\%$ (left panel) of low mass group R halos have lost more than $5\%$ of their peak mass at $z = 0$; however, nearly $70\%$ (right panel) of these same halos experienced a peak mass loss of more than $5\%$ since their last major merger.  This points out that mass loss following a major merger is the norm, rather than the exception.  Further, about $50\%$ of all group R halos experienced a peak mass loss of between 5 and $15\%$ (about $20\%$ lost more than $15\%$ of their peak mass, and $30\%$ lost less than $5\%$). Most of these halos have resumed accreting material by $z = 0$, substantially reducing the amount of mass loss present in the $z = 0$ distribution.  While high mass halos lose less mass from mergers than low mass halos, there is not a strong mass dependence among group R halos.  Group N halos still exhibit weak amounts of mass loss, but significantly less than the key TS and R mass loss groups.  About $12\%$ of group N halos experience a peak mass loss of greater than $5\%$, with very little mass dependence.  These group N distributions seem roughly consistent with the amount of mass loss we'd expect from minor mergers alone, which we expect to be the leading cause of mass loss among group N halos.}
    \label{fig:bsr_group_cdf}
\end{figure*}

%
%

\begin{figure*}
	\centering
	\includegraphics[angle=0,trim=100 40 6 15, clip,width=\textwidth]{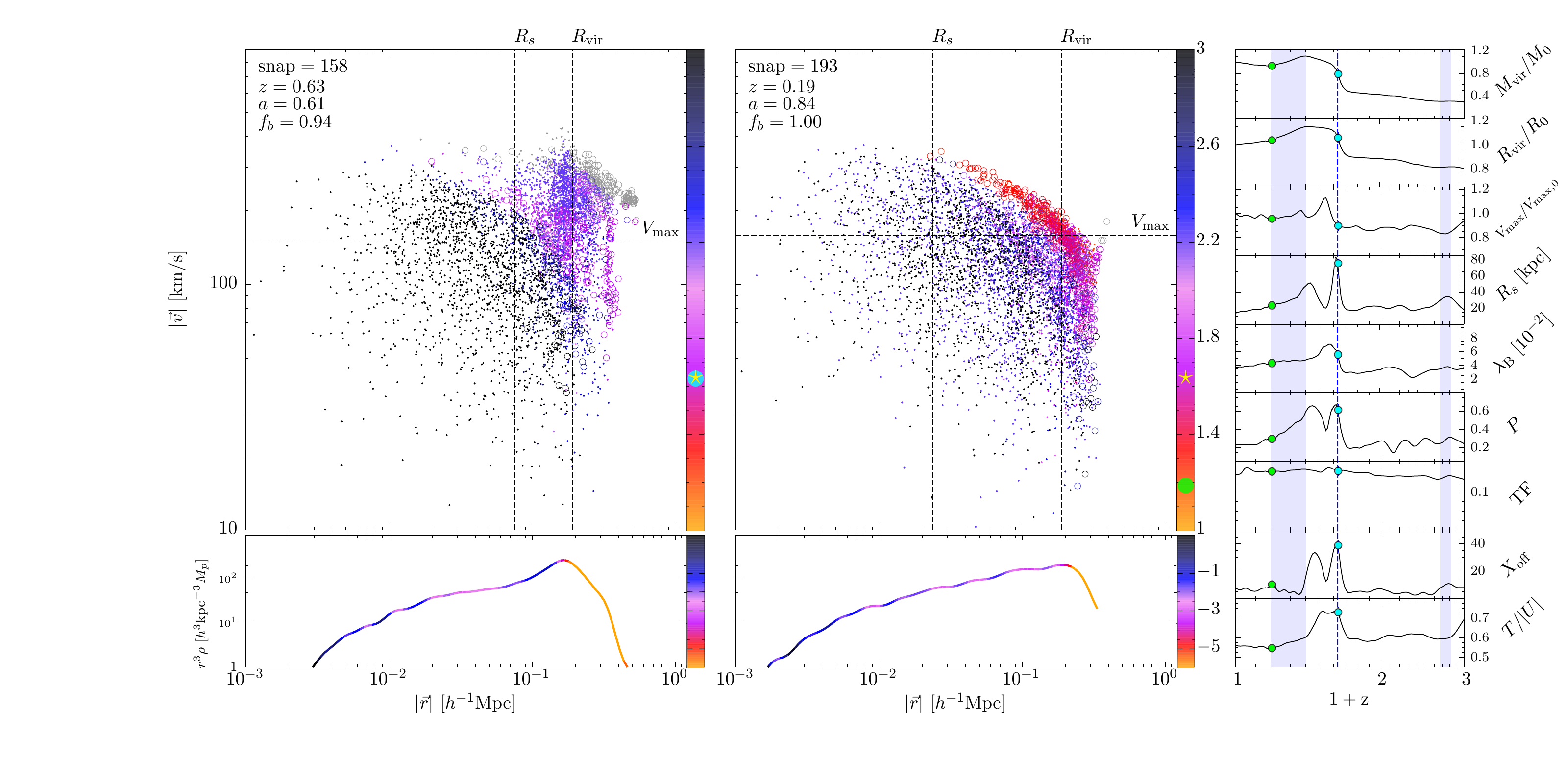}
    \caption{Particle distributions, density profiles, and halo property evolution of an individual group R halo at two distinct snapshots.  Large panels show the distribution of particles at $z = 0.63$ (left) and $z = 0.19$ (right). On the x-axis (y-axis) we plot the magnitude of the radial (velocity) vector of each particle, with the center and bulk velocity of the halo as the origin. Individual particles are coloured according to the redshift they are first accreted onto the halo (the associated colour bar is in units of $1+z$). Solid points represent particles that will remain bound to the halo at $z = 0$, while open circles represent particles that will be removed from the halo by $z = 0$.  Gray coloured particles are unbound.  In each large panel, we also indicate the current simulation snapshot, scale factor, fraction of particles that are bound ($f_{b}$), and the current location of $\vmax$, $\rs$, and $\rvir$.  In the associated colour bar on the right, the yellow star indicates the occurrence of a major merger, while the cyan and green circles indicate the current time. The (bottom) panels below the particle distributions show the halo density profiles at each snapshot.  We scale the profiles by $r^3$ to highlight deviations of the outer profile from a $r^{-3}$ (NFW) slope.  The associated color bar on the right indicates the $\log r - \log \rho$ slope of the density profile.  For an NFW halo, the slope would be -1 (dark blue) below $\rs$, and -3 (purple) 
    above $\rs$.  Black sections indicate extremely shallow profiles, while red and orange sections indicate very steep profiles.  The rightmost column of plots show the full evolution of mass, virial radius, $\vmax$, scale radius, spin parameter, prolateness, tidal force, $\xoff$, and virial ratio for this halo from $z = 2$ to $0$.  The green (cyan) circles indicate the location of the left (right) main panels.  The vertical blue dashed line shows where a major merger occurs, and the blue shaded regions indicate periods of mass loss. Mass, virial radius, and $\vmax$ are each normalized by their $z = 0$ values. We've chosen these two snapshots to highlight the way a major merger followed by mass loss affects halo properties.  The left main panel ($z = 0.63$) coincides with the last major merger.  We see that a large cluster of recently accreted blue and pink material has just crossed the virial radius -- this is the merging halo.  The presence of this new material temporarily reduces the slope of the outer density profile, causing the scale radius to expand (along with $\xoff$, $P$, and $\tu$). The right main panel ($z = 0.19$) coincides with the time of peak net mass loss (minimum $\mvir$ since $\mpeak$) following the major merger.  At this point the halo has revirialized and its density profile is closer to NFW.  Some new material continues to flow in along the outermost layer of the distribution, while material from the merger still flows out in the opposite direction.  Much of the material accreted the newly accreted material has settled in this outer band, with very few particles occupying the low $v$, low $r$ region of the distribution.  Some halo properties like $\rs$, $\lambdaB$, and $P$ still remain elevated compared to their pre-merger levels, but these continue to decline as the accretion rate increases.}
    \label{fig:PD_R}
\end{figure*}

\subsection{Post-Merger Relaxation and Mass Loss}
\label{sec:MM}

The most common mass loss mechanism for dark matter halos is the shedding of excess high energy material in the aftermath of a halo merger event. From Fig. \ref{fig:mechanism_freq} we can see that halos with a recent major merger but no recent history of high tidal force (group R) are by far the most common group of mass loss candidates. In Fig. \ref{fig:bsr_group_cdf}, we show the cumulative distribution functions of the mass loss ratio for all halos in groups N, TS, and R to highlight the distribution of diminished halos among those that, respectively, do not appear to be influenced by tidal or post-merger mass loss, are susceptible to tidal stripping exclusively, and are susceptible to post-merger mass loss exclusively. We additionally plot in the right column the minimum mass loss fraction since the peak mass for halos in groups N and TS and since the scale factor of the last major merger for halos in group R. We see that about $6\%$ (low mass) -- $8\%$ (high mass) of group N halos have lost more than $5\%$ of their peak mass at $z = 0$, a much lower fraction than the $12-22\%$ of all halos (Fig. \ref{fig:stripping_cdf}). Essentially no group N halos have lost more than $20\%$ of their peak mass. Clearly, removing halos associated with possible mass loss mechanisms (groups TS, R, and TS+R) also removes the majority of diminished halos, leaving only a small percentage of weakly diminished halos. The small number of remaining group N halos experienced mass loss either due to a third mechanism, or due to weaker manifestations of tidal stripping (with tidal force below 1) or merger-induced mass loss (from a minor merger); we will further investigate these possibilities shortly.

Looking at the distribution of the mass loss fraction among group R halos (those that experienced a recent major merger but no tidal stripping; Fig. \ref{fig:bsr_group_cdf}, middle row), we see that roughly $18\%$ (low mass) -- $32\%$ (high mass) of halos are diminished.  However, the distribution changes considerably when looking at the minimum mass loss fraction since the last major merger; in this case, the fraction of diminished halos jumps to roughly $48-68\%$ of all group R halos.  Furthermore, we see that nearly all group R halos experienced mass loss of a few percent of their peak mass or more, with lower mass halos tending to lose more mass than higher mass halos.  Most major mergers cause losses of around $5-15\% \ \mpeak$ and very few result in heavy mass loss of more than $\sim 20\% \ \mpeak$, consistent with the trends from Fig. \ref{fig:stripping_properties_correlation} that peak around $\mvir / \mpeak \approx 0.90$ for low mass halos (lower concentration, higher spin, more prolate halos). Once halos have relaxed and begun re-accreting material after a major merger, they will eventually return to and exceed their original peak mass and restore the mass loss fraction to 1.  This explains why the left panel (mass loss fractions at $z = 0$) is so different from the right panel; many of the halos have completed the mass loss process and begun re-accreting, reducing their apparent mass loss fraction.

\begin{figure*}
	\centering
	\includegraphics[angle=0,trim=100 40 6 15, clip,width=\textwidth]{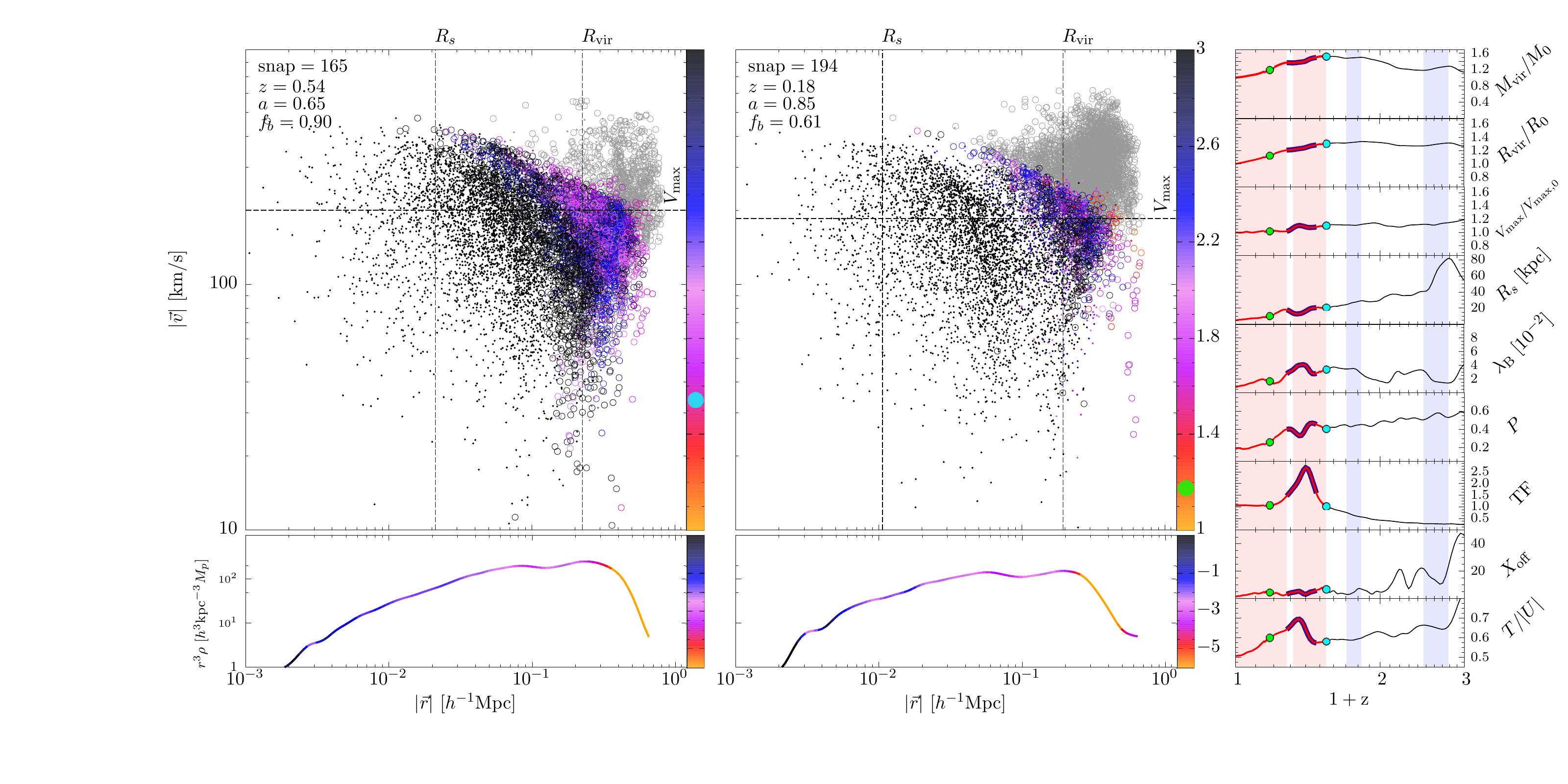}
    \caption{Same as Fig. \ref{fig:PD_R}, but showing a group TS halo at $z = 0.54$ and $z = 0.18$.  In this case, no major mergers occur.  In the property evolution plots in the rightmost column, red shading indicates periods of mass loss where tidal force is greater than 1 (i.e. tidal stripping).  Blue shading indicates periods of mass loss where tidal force is less than 1.  The bold red portion (from roughly $z = 0.54$ to $0$) indicates where tidal force is greater than 1, and the additional bold purple section shows when this halo is a subhalo of a more massive halo.  We've chosen these two snapshots to highlight the affect of tidal stripping on halo properties.  The left main panel ($z = 0.54$) coincides with the peak mass and where tidal force becomes greater than 1.  We see that most of the (relatively recently accreted) material in the outer part of the halo will be lost by $z = 0$.  Shortly after this, the halo becomes a subhalo and experiences much stronger tidal forces.  The right main panel ($z = 0.18$) coincides with the minimum tidal force (roughly $\tf = 1$) after the halo again becomes a distinct halo.  At this point halo mass, virial radius, $\vmax$, scale radius, spin, and prolateness have all decreased relative to their pre-subhalo values.  We can see a clear separation between the distribution of material in the inner halo and outer halo, and a corresponding steepening of the density profile around this same radius.  The decrease in scale radius is a direct consequence of this steepening of the outer density profile. By $z = 0$, the halo will be nearly entirely stripped of the puffed-out outer region.}
    \label{fig:PD_TS}
\end{figure*}

{\color{black}Let's look in detail at a typical example of mass loss following a major merger.  In Fig. \ref{fig:PD_R}, we show for an individual halo the particle distribution and density profile at two different snapshots ($z = 0.19$ and $z = 0.63$), as well as the full evolution of various halo properties since $z = 2$, including halo mass ($\mvir$), virial radius ($\rvir$), maximum circular velocity ($\vmax$), NFW scale radius ($\rs$), spin parameter ($\lambdap$), prolateness (P), tidal force (TF), offset between density peak and center of mass ($\xoff$) and virial ratio ($\tu$). We've chosen these two snapshots to highlight the effects of this mass loss scenario on halo properties. At $z = 0.63$ (left main panel), the merging halo has just crossed the virial radius of the main halo; this is the snapshot that \rockstar\ associates with the merger event. At the same time, we see a sharp increase in halo mass, virial radius, scale radius, spin parameter, prolateness, $\xoff$, and virial ratio.  $\vmax$ displays a slightly delayed response, and tidal force does not change appreciably.  The density profile develops a distinct hump due to the merging core that migrates towards the center of the main halo before splashing back.  This behaviour causes corresponding oscillations in the scale radius (which fluctuates due to a poor NFW fit to the temporarily shallower-than-expected outer profile), prolateness, and $\xoff$.  We don't expect to see oscillations in all properties; spin parameter, for example, is not strongly affected by these dynamics.  The halo quickly begins to relax, even as additional material flows in following the main merger.  By the time the halo mass and virial radius have peaked, most other properties are settling back to historically typical values.  The net mass accretion rate turns negative and the halo mass and virial radius start shrinking as high energy loosely bound material escapes beyond the virial radius.  At $z = 0.19$ (right main panel), the net accretion rate turns positive again; this is the minimum value of $\mvir/\mpeak$ this halo will reach due to this merger ($\sim 0.8$ in this case).  At this point, most halo properties have settled considerably, but scale radius, prolateness, and spin parameter are still somewhat elevated above typical levels.  All of the recently accreted material has accumulated along the outermost curved layer of the distribution, with material continuing to flow both in and out along this trajectory.  The density profile has smoothed out and appears roughly NFW.  We have also created movies that show the full evolution of the halos in Figs. \ref{fig:PD_R} and \ref{fig:PD_TS}, which are publicly available. \footnote{\url{https://goo.gl/Qtf9i9}}}

\begin{figure*}
	\centering
	\includegraphics[trim=20 10 60 2, clip, width=0.7\textwidth]{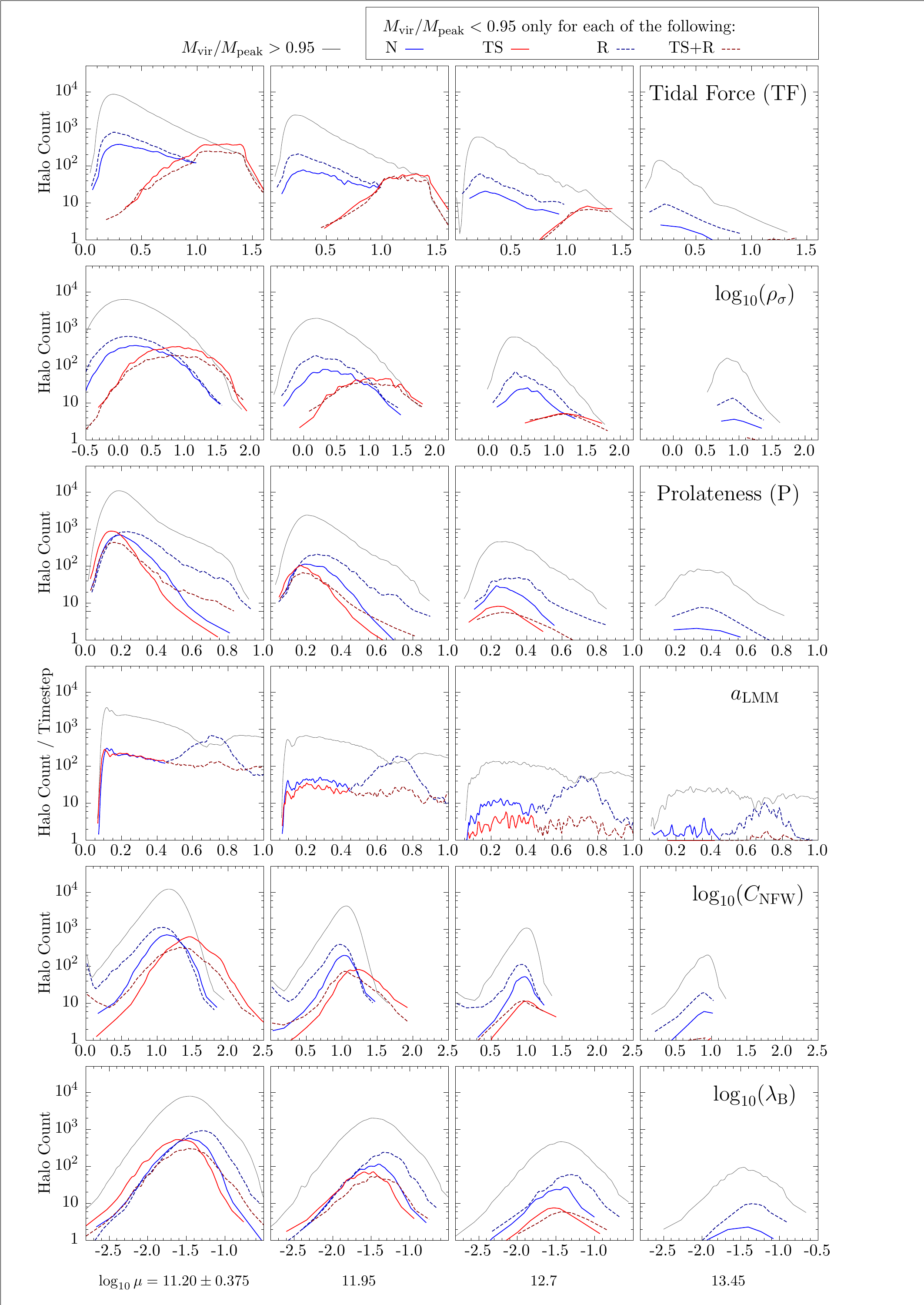}
    \caption{Distributions of tidal force, local density, prolateness, scale factor of last major merger, NFW concentration, and Bullock spin parameter for all distinct halos that have not lost a significant amount of mass and for distinct halos that have lost mass in Groups N, TS, R, and TS+R. {\color{black}We use the same mass bin definitions as in Fig. \ref{fig:stripping_cdf}.} Vertical axes show the number of halos in each horizontal bin; all distributions in a given row have the same horizontal bin width to allow for direct comparison to other distributions in the same panel.  Values on the horizontal axes correspond to the halo property labelled in the rightmost column of a given row.  We divide these distributions into our usual four mass bins, with the lowest mass bin in the leftmost column.  The grey lines indicate distributions for all halos in a given mass bin that have not lost more than $5\%$ of their mass since $\mpeak$.  The solid blue, solid red, dashed blue, and dashed red lines represent halos that have lost more than $5\%$ of their mass since $\mpeak$ and additionally belong to groups N, TS, R, and TS+R, respectively.  We see that most halos experiencing high tidal forces and in high density regions have been tidally stripped by more than $5\%$, while diminished halos in groups N and R have similar tidal force and local density distributions compared to undiminished halos.  Halos experiencing purely tidal stripping (group TS) are less prolate, more concentrated, and have lower spins than undiminished halos.  Those that experienced purely merger induced mass loss (group R) are more prolate, less concentrated, and have higher spins.  The majority of halos with a last major merger around $a = 0.7$ are diminished group R halos. }
    \label{fig:pdf_mechanisms1}
\end{figure*}

\begin{figure*}
	\centering
	\includegraphics[trim=10 60 55 55, clip, width=\textwidth]{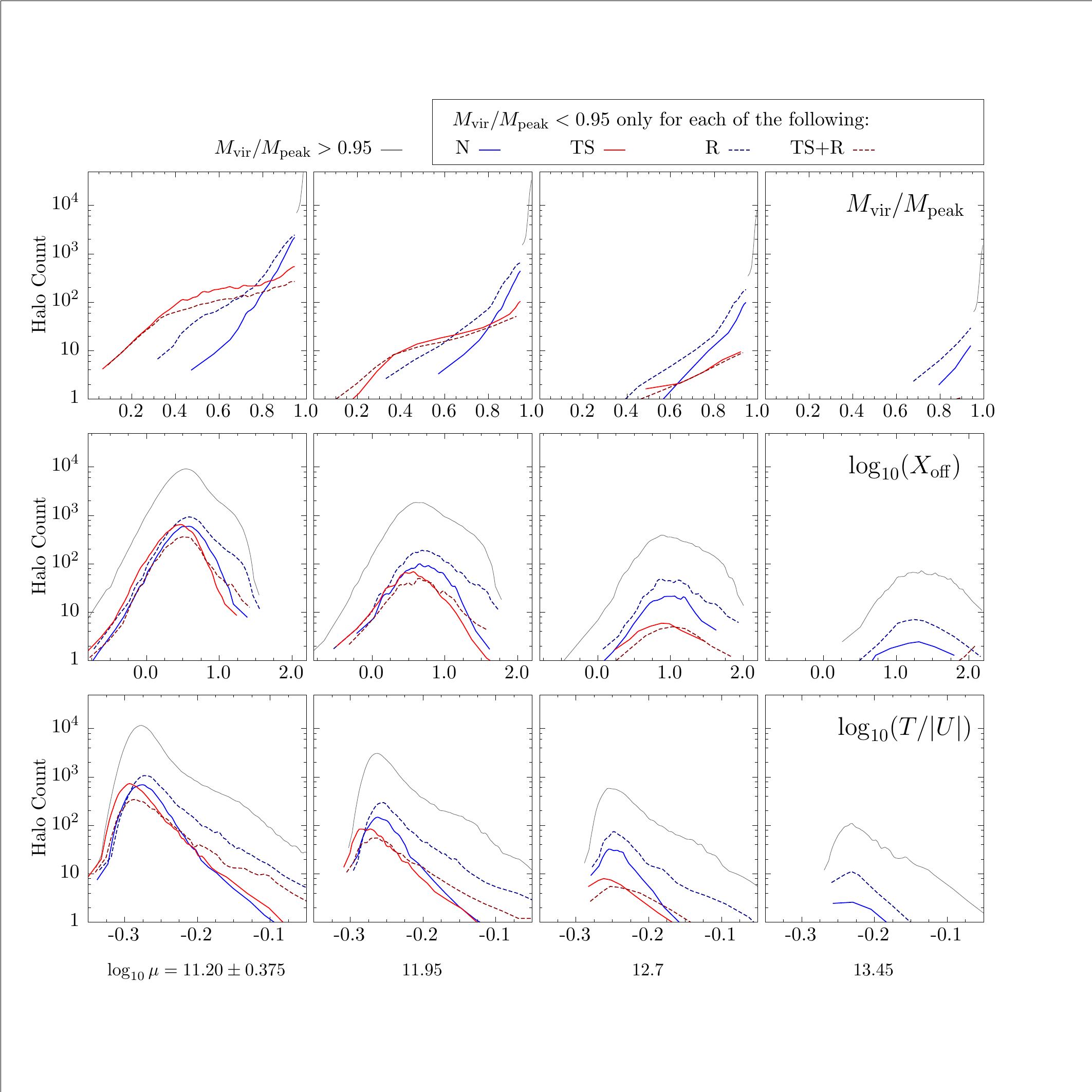}
    \caption{Same as Fig. \ref{fig:pdf_mechanisms1}, but showing mass loss ratio ($\mvir/\mpeak$), offset between halo center of mass and density peak ($\xoff$), and virial ratio ($\tu$). {\color{black}We use the same mass bin definitions as in Fig. \ref{fig:stripping_cdf}.} We see that Group R halos dominate the population of halos that have only experienced moderate mass loss (between 5 and $20\%$ of their peak mass for low mass halos), while tidally stripped Group TS halos are the most common among heavily diminished halos.  Group TS halos have slightly lower $\xoff$ and $\tu$ on average compared to undiminshed halos, while Group R halos have slightly higher $\xoff$ and $\tu$ on average.  These trends are consistent with our proposed mass loss mechanisms of tidal stripping and post-merger relaxation for Groups TS and R, respectively.}
    \label{fig:pdf_mechanisms2}
\end{figure*}

In Figs. \ref{fig:pdf_mechanisms1} and \ref{fig:pdf_mechanisms2}, we show the distributions of many halo properties for diminished halos from groups N, TS, R, and TS+R along with non-diminished halos from all groups.  We select only diminshed halos from each group to highlight how different types of mass loss influence other halo properties.  The distributions are not normalized and have consistent bin sizes within each panel, providing an accurate representation of the relative abundances of halos in each panel. Starting with tidal force (TF; Fig. \ref{fig:pdf_mechanisms1} Row 1), we see that most halos with very high TF values ($>1$) are diminished group TS halos.  Note that this indicates that diminished group TS halos are more abundant than non-diminished TS halos (which would contribute to the gray line), consistent with the CDFs from Fig. \ref{fig:bsr_group_cdf}.  We also see a tail of diminished group TS and TS+R halos with $\tf < 1$ at $z = 0$, which indicates that at least some halos experience high TF shortly after they reach their peak mass (qualifying them to be in group TS or TS+R), but subsequently return to lower TF regions by $z = 0$. The TF distribution of diminished group N and R halos is similar to that of non-diminished halos, and is limited to the range $0 < \tf < 1$ by the group assignment criteria.  The distributions of local density (Fig. \ref{fig:pdf_mechanisms1} Row 2) provide qualitatively similar information as the TF distributions \citep[since TF correlates strongly with local density; see][]{Lee17}.  However, we do observe more overlap between local density distributions of low TF (groups N and R) and high TF (groups TS and TS+R) halos than for the TF distributions.  

Especially notable for halos that recently experienced a major merger are the next several properties from Fig. \ref{fig:pdf_mechanisms1}: P, $\almm$, $\cnfw$, and $\lambdaP$.  Among the diminished halos, we see that halos from group R are significantly more prolate that halos from groups N, TS, and TS+R on average.   Group R halos have a peak $\cnfw$ similar to that of group N halos, but have notably more low concentration halos than diminished halos from the other groups, and also have the highest spin parameters.  The distributions of $\almm$ reflect the group selection criteria ($\almm = 0.45$ is the dividing epoch between groups N/TS and R/TS+R), but also reveal an interesting peak around $\almm = 0.7$ for diminished group R halos.  This peak provides an indication of the characteristic mass loss timescale -- the mean time delay between a major merger event and subsequent peak mass loss (in this case fixed at $z = 0$).  Halos in group R with $\almm < 0.7$ likely have begun accreting again, but may not yet have surpassed their previous peak mass by $z = 0$, while group R halos with $\almm > 0.7$ are likely actively losing mass and have not yet reached their minimum mass.

From Fig. \ref{fig:pdf_mechanisms2} Row 1, we see that weakly diminished halos (with a mass loss fraction $\sim 0.8-0.95$) are dominated by group R, followed by group N.  Strongly diminished group R halos are much less common than those from groups TS and TS+R, reflecting the same trends observed from the CDFs from Fig. \ref{fig:stripping_cdf}.  The distribution of mass loss among diminished group N halos is similar in shape to that of diminished group R halos, but falling off more quickly with increasing mass loss.  {\color{black}As do group R halos, weakly diminished group N halos ($\mvir/\mpeak \grtsim 0.85$) dominate over group TS and TS+R halos in the same mass loss regime, but group N halos are the least common among those that have lost more than $20\%$ of their peak mass}. This strongly suggests that the processes responsible for mass loss in group N halos may be similar to the relaxation-based process attributed to mass loss in group R halos (e.g. mass loss after a minor merger).  We additionally see in the remaining rows of Fig. \ref{fig:pdf_mechanisms2} that diminished group R halos are on average the least symmetric (highest $\xoff$), and the least relaxed (highest $\tu$) compared to diminished halos from the other groups.  This is consistent with the the proposed major-merger induced mass loss mechanism.  Halos undergoing a major merger would become temporarily unrelaxed and have two cores, increasing both $\tu$ and $\xoff$ before simultaneously undergoing mass loss and relaxation.

We've additionally followed the most massive progenitors of the same populations plotted in Figs. \ref{fig:pdf_mechanisms1} and \ref{fig:pdf_mechanisms2} (diminished group N, R, TS, TS+R and not diminished halos) and determined median halo properties of these populations out to high redshift ($z \sim 7$).  In Fig. \ref{fig:stripping_prog_hist_1}, we follow the evolution of NFW concentration ($\cnfw$), Bullock spin parameter ($\lambdaB$), virial mass ($\mvir$), and specific mass accretion rate ($\mar$).  We find that diminished group R halos all exhibit a significant change in behaviour around $z \approx 0.5$ ($a \approx 0.7$), consistent with the strong peak in $\almm \approx 0.7$ from Fig. \ref{fig:pdf_mechanisms1} Row 4.  At this epoch, diminished group R halos undergo a decrease in concentration, sharp increase in spin, and increase in specific accretion rate.  These halos subsequently regress back towards higher concentrations and lower spins as they approach $z = 0$, but still remain with significantly lower concentrations and higher spins at $z = 0$ compared to the other populations plotted.  After the increase in accretion rate coincident with the last major merger, these halos experience a sharp drop in accretion rate than dips below zero briefly, before returning to zero at $z = 0$.  This clearly illustrates the mass loss phenomenon these recently merged halos experience -- an initial surge in mass from the merging halo (which also sets $\mpeak$), followed by a period of mild mass loss as the halo relaxes and sheds high energy material.  Interestingly, diminished group N halos exhibit each of these trends as well, but with less deviation from their historic median values compared to group R halos. This gives further weight to the suggestion that diminished group N halos have likely experienced mass loss due to a minor merger, rather than unaccounted-for tidal stripping or some other new mass loss mechanism.  We additionally note that all halo populations plotted have similar median concentrations and spins at high redshift, while diminished group R halos were the least massive and slowest accreting halos, followed by group N, group TS+R, and group TS halos, respectively.  These trends are consistent across all mass bins shown, a further indication that this merger-induced mass loss is a general phenomenon affecting a wide range of halo masses in a similar fashion.
  
From Fig. \ref{fig:stripping_prog_hist_2} Rows 2 and 3, we see that for diminished group R halos, changes in concentration ($\cnfw \equiv \rvir / \rs$) stem almost entirely from changes in $\rs$ rather than $\rvir$.  The scale radius is strongly amplified during the major merger, and remains substantially elevated at $z = 0$ even after regressing during the post-merger relaxation phase.  We do not interpret this dramatic fluctuation in scale radius as a shift in the $\log{\rho} - \log{\rvir}$ slope change from -1 to -3 as is typically expected for an NFW halo, but rather as a result of a poor fit to a halo profile that no longer resembles the NFW profile.  Attempting to fit an NFW profile to an unrelaxed halo with a recent accumulation of mass in the outer regions will tend to produce an artificially elevated scale radius in an attempt to compensate for the shallower than expected outer profile slope.  Again, we note that diminished group N halos mirror the $\rs$ evolution of diminished group R halos, but do not experience as dramatic a response.

Finally, we see from Fig. \ref{fig:stripping_prog_hist_3} that diminished group R halos undergo a sharp increase in prolateness, asymmetry, and virial ratio coincident with the typical last major merger epoch.  The prolateness measured at $\rvir$ peaks strongly shortly after the merger and then subsides, but remains elevated compared to all other groups at $z = 0$.  When measured at $\rfive$, prolateness peaks shortly after $\pvir$ does, and is less dramatic than the peak in $\pvir$, consistent with the picture that a typical merging halo will initially disrupt the outer halo and become substantially stripped before punching into the inner part of the halo.  The peak in $\doff$ and $\tu$ coincide with the peak in $\pvir$, indicating that halos become less symmetric (halo density peak and center of mass becoming misaligned to due the addition of an off-center clump of material from the merging halo) and unrelaxed.  Of  
the trends shown in Fig. \ref{fig:stripping_prog_hist_3}, $\tu$ shows the least dramatic divergence for diminished group R halos compared to diminished halos from groups TS, TS+R, and N.  By $z = 0$, diminished group R halos have largely completed the relaxation process, returning to values of $\doff$ and $\tu$ comparable to the non-diminished halos.  Group N diminished halos exhibit similar behavior as those from group R, but are much more weakly elongated, especially when measured at $\rfive$, and typically fully relaxed by $z = 0$.  These trends are consistent for all four mass bins shown, albeit with more uncertainty present at higher masses due to low number statistics.

This leaves us with a detailed and coherent picture of how major mergers affect the properties of dark matter halos and ultimately induce mass loss through relaxation.  Diminished group R halos, which are those we identify at $a=1$ 
as being most directly subject to this process, typically experienced a major merger around $a = 0.7$.  The physical properties of the host halo begin to change dramatically as the incoming halo impinges on the host, becomes tidally disrupted, and deposits material on the host halo.  The host halo becomes less concentrated, since the NFW scale radius increases to accommodate the recent accumulation of material in the outer profile.  Simultaneously, the spin parameter and halo prolateness typically increase due to a non-zero impact parameter and a preferred axis of accretion (mergers would tend to flow along the directions of pre-existing filaments or sheets).  The host halo becomes increasingly unrelaxed and less symmetric during this brief mass accretion phase of the merger.  As the core of the merging halo becomes fully disrupted and integrated into the host halo, the system relaxes and once again becomes more concentrated, lower spin, less prolate, and more symmetric.  The net mass accretion rate turns negative {\color{black}as the amount of newly infalling material decreases and high energy material gradually dissipates or is moved 
outside the virial radius of the halo.}  This is what constitutes the post-merger mass loss mechanism.  Post-merger mass loss peaks very shortly before $z = 0$ for diminished group R halos, after the halos have relaxed considerably, but still have considerably higher spin parameters than before the merger.




%



\subsection{Tidal Stripping}
\label{sec:TS}

The other principal mass loss mechanism we've identified in this work is the tidal stripping of material from the outer regions of halos, which primarily occurs in high density regions as a result of strong tidal forces from a nearby massive halo.  In Fig. \ref{fig:bsr_group_cdf} Row 1, we see that nearly all group TS halos (those that have had $\tf > 1$ at any point since their peak mass), have less mass at $z = 0$ compared to their peak mass.  A majority ($66\%$) of halos in the lowest mass bin ($\log_{10} \mu = 11.2 \pm 0.375$) are diminished at $z = 0$, and about $30\%$ of these halos have lost more than $30\%$ of their mass, a far greater fraction than group R halos from the same mass bin, even when comparing to their maximum mass loss since their last major merger.  The left and right panels of the group TS distributions are nearly identical, indicating that these halos are at their historic minimum mass at $z = 0$.  In other words, halos that undergo tidal stripping do not generally recover and begin accreting rapidly again.  If that were common behaviour, we would expect to see a vertical shift in the right panel (minimum $\mvir/\mpeak$ since $\mpeak$).  This suggests that most group TS halos either become subhalos or disappear completely in subsequent timesteps.  There is also a strong mass dependence in the distribution of mass loss fraction among group TS halos.  High mass halos are much less likely to be heavily stripped than low mass halos ($66\%$ of low mass halos are diminished at $z = 0$, compared to only $24\%$ of high mass halos).  This may be a result of fewer sufficiently massive neighbouring halos for high mass halos than low mass halos.  High mass halos tend to be dispersed throughout the cosmic web in nodes and thick filaments, and are less likely to come in contact with another halo sufficiently massive to tidally disrupt it and induce mass loss.  Additionally, we have not excluded the possibility of recent minor mergers contributing to mass loss.  For low mass group TS halos, we expect mass loss to be primarily due to tidal stripping, but some of the high mass diminished group TS halos in our sample may be experiencing mass loss due to a recent minor merger; indeed, the distribution of of these halos is similar in shape to the distribution of group R and group N halos. 

{\color{black}Again, we take a look at an example of an individual halo experiencing significant mass loss via tidal stripping.  Fig. \ref{fig:PD_TS} shows the particle distribution and density profile of the halo at two key snapshots ($z = 0.54$ and $z = 0.18$), as well as the full evolution of many halo properties since $z = 2$.  We've chosen these two snapshots to highlight the effects of this mass loss scenario on halo properties. At $z = 0.54$ (left main panel), the halo is just crossing into a region with tidal force > 1, indicating that the Hill radius $R_{\mathrm{Hill}}$ from a massive nearby halo is smaller than the virial radius of the halo (i.e., that weakly bound material in the outer part of the halo will be stripped away).  This snapshot also coincides with the peak mass of the halo and the start of nearly continuous mass loss for the remainder of the simulation runtime.  At $\mpeak$, most other halo properties have not yet changed significantly in response to tidal effects, although we already see a mild steepening of the outer density profile just before $\rvir$, as well as a build up of bound and unbound material outside the virial radius.  Shortly after $\mpeak$, the halo punches into a more massive halo and becomes a subhalo (indicated by the thick purple line segment in the property evolution panels on the right).  This results in a brief increase in virial ratio and a strong peak in tidal force; most other halo properties fluctuate mildly during this period, though we also expect halo finder noise to increase significantly in such regions of extreme density.  The (sub)halo re-emerges from the massive halo and again becomes a distinct halo.  At $z = 0.18$ (right main panel) the tidal force drops to 1 again, marking the end of the very high tidal force event, although it 
does not decrease below 1 for the remainder of the simulation.  By this time, halo mass and radius have decreased significantly compared to their peak values, and continue to decline.  Scale radius, spin parameter and shape have all decreased notably.  The outer density profile falls off faster than $r^{-3}$ just before $\rvir$, causing the NFW fit to artificially suppress the scale radius.  It is perhaps clearer to visually detect from the particle distribution the separation between the inner core of the halo and the soon-to-be-stripped mass of material sitting near the virial radius.  Likely, the removal of so much material on the outskirts of the halo, in particular high energy material on very elliptical orbits, contributes to the overall sphericalization and decrease in spin parameter that we see during this process.  These trends largely continue towards $z = 0$, by which time the halo has lost roughly $35\%$ of its peak mass.

The trends observed for the individual halo in Fig. \ref{fig:PD_TS} are consistent with those observed for the whole population of group TS halos.} From Fig. \ref{fig:pdf_mechanisms1}, we'll focus our discussion on Rows 3-6.  We see that diminished group TS halos are the least prolate (most spherical), compared to all other populations plotted.  At least at low masses, the prolateness distribution of group TS halos both peaks at a lower prolateness and is skewed towards lower prolateness values.  Group TS halos have significantly higher average concentrations compared to the other groups.  The concentration distribution peaks around $\cnfw = 32$ for group TS (compared to only about $11$ for group R halos), but also has a higher dispersion.  Very few diminished group TS halos have concentrations below about 10, while this is common for group R and non-diminished halos.  These halos also have the lowest average spin parameter compared to the other groups plotted.  While the peak spin parameter of diminished group TS halos is only slightly below that of group N halos, the distribution is skewed towards lower spins.  These trends indicate that diminished group TS halos are most differentiated from other halos by their much higher average concentration, but also are typically more spherical (less prolate) and have lower spins.  These trends are all weaker at higher masses; the properties of group TS halos are more differentiated from the remaining populations at lower masses.

As expected, diminished group TS+R halos display trends that are somewhat intermediary between those of group TS and group R.  They have a prolateness distribution similar to that of the non-diminished halos, though peak at slightly lower prolateness.  There is a subtle peak in $\almm$ around $0.7$, coincident with the peak in group R halos, indicating merger-induced mass loss is at least partly responsible for the differences in halo properties of group TS+R halos compared to non-diminished halos.  These halos also have a higher average concentration than group R, N, and non-diminished halos, but less than group TS halos.  The spin distribution of group TS+R halos is similar in shape to that of group N and non-diminished halos, but with an excess of high spin halos (at least for the lowest mass bin).

\begin{figure*}
	\centering
	\includegraphics[trim=16 154 60 174, clip, width=\textwidth]{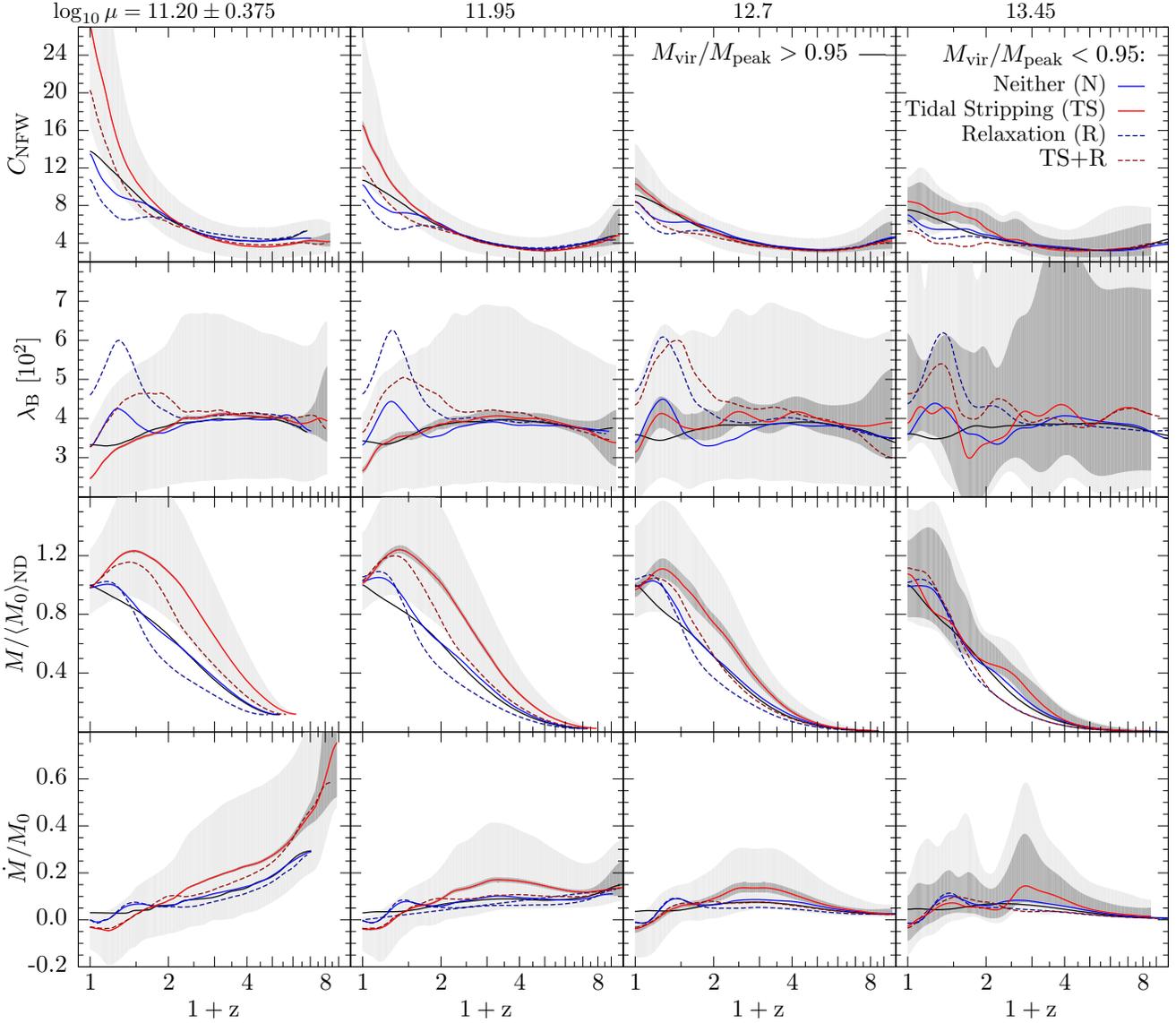}
    \caption{NFW Concentration, spin parameter, mass, and specific mass accretion rate histories for all distinct halos that have not lost a significant amount of mass (not diminished (ND): $\mvir/\mpeak > 0.95$) and for distinct halos that have lost mass ($\mvir/\mpeak < 0.95$) in Groups N, TS, R, and TS+R. {\color{black}We use the same mass bin definitions as in Fig. \ref{fig:stripping_cdf}.} The curves reflect median properties of the progenitors of the z = 0 halo populations. The dark grey shading reflects the 95\% confidence interval on the median and the light grey shading reflects the $20-80\%$ dispersion of each property, shown only for Group TS halos.  Each halo mass curve is normalized to the $z = 0$ value of the not diminished (ND) curve in each mass bin.  We see that halos experiencing purely tidal mass loss (Group TS) experience amplified concentrations, reduced spin parameters, and heavily reduced accretion rates and halo mass at late times.  In contrast, halos that underwent purely merger induced mass loss (Group R) exhibit temporarily reduced concentrations, strongly amplified spins, and a recent burst of accretion, followed by mild mass loss.  Group TS+R halos display milder trends consistent with both tidal stripping and relaxation, while Group N halos display trends that are most consistent with those from Group R, but subdued, suggesting these may be halos experiencing weaker relaxation-based mass loss from minor mergers.} 
    \label{fig:stripping_prog_hist_1}
\end{figure*}

\begin{figure*}
	\centering
	\includegraphics[trim=16 154 60 174, clip, width=\textwidth]{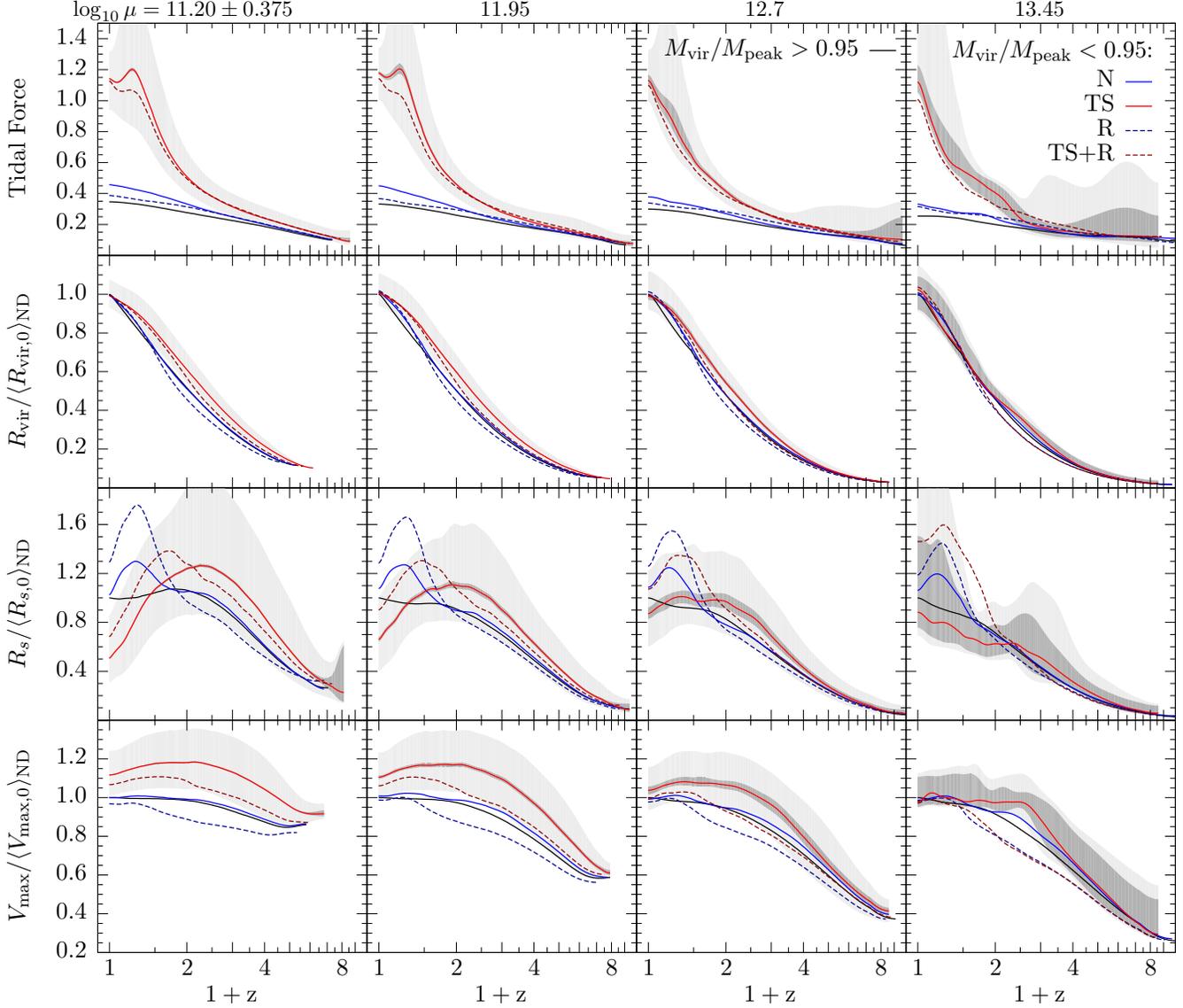}
    \caption{Same as Fig. \ref{fig:stripping_prog_hist_1}, but showing tidal force, virial radius ($\rvir$), scale radius ($\rs$), and maximum circular velocity ($\vmax$).  To efficiently compare different mass bins, we normalize $\rvir, \rs,$ and $\vmax$ by the median values of the not-diminished (ND) population at $z = 0$.  Note that the $\rvir$ curves must converge at $z = 0$ as a consequence of the halo mass curves (Fig \ref{fig:stripping_prog_hist_1} Row 3) converging at $z = 0$.  We see that halos experiencing purely tidal mass loss (Group TS) typically experience much stronger tidal forces starting around $z \approx 1-2$ and peaking shortly before $z = 0$, strongly depressed scale radii and mildly reduced maximum circular velocity, both roughly coincident with the increase in tidal force.  In contrast, halos that underwent purely major merger induced mass loss (Group R) exhibit consistently low tidal force, temporarily amplified scale radii, and a jump in maximum circular velocity roughly coincident with the major merger.  Group TS+R halos display milder trends consistent with both tidal stripping and post-merger mass loss, while Group N halos display trends that are most consistent with those from Group R, but subdued, suggesting these may be halos experiencing weaker relaxation-based mass loss from minor mergers.}
    \label{fig:stripping_prog_hist_2}
\end{figure*}

\begin{figure*}
	\centering
	\includegraphics[trim=16 154 60 174, clip, width=\textwidth]{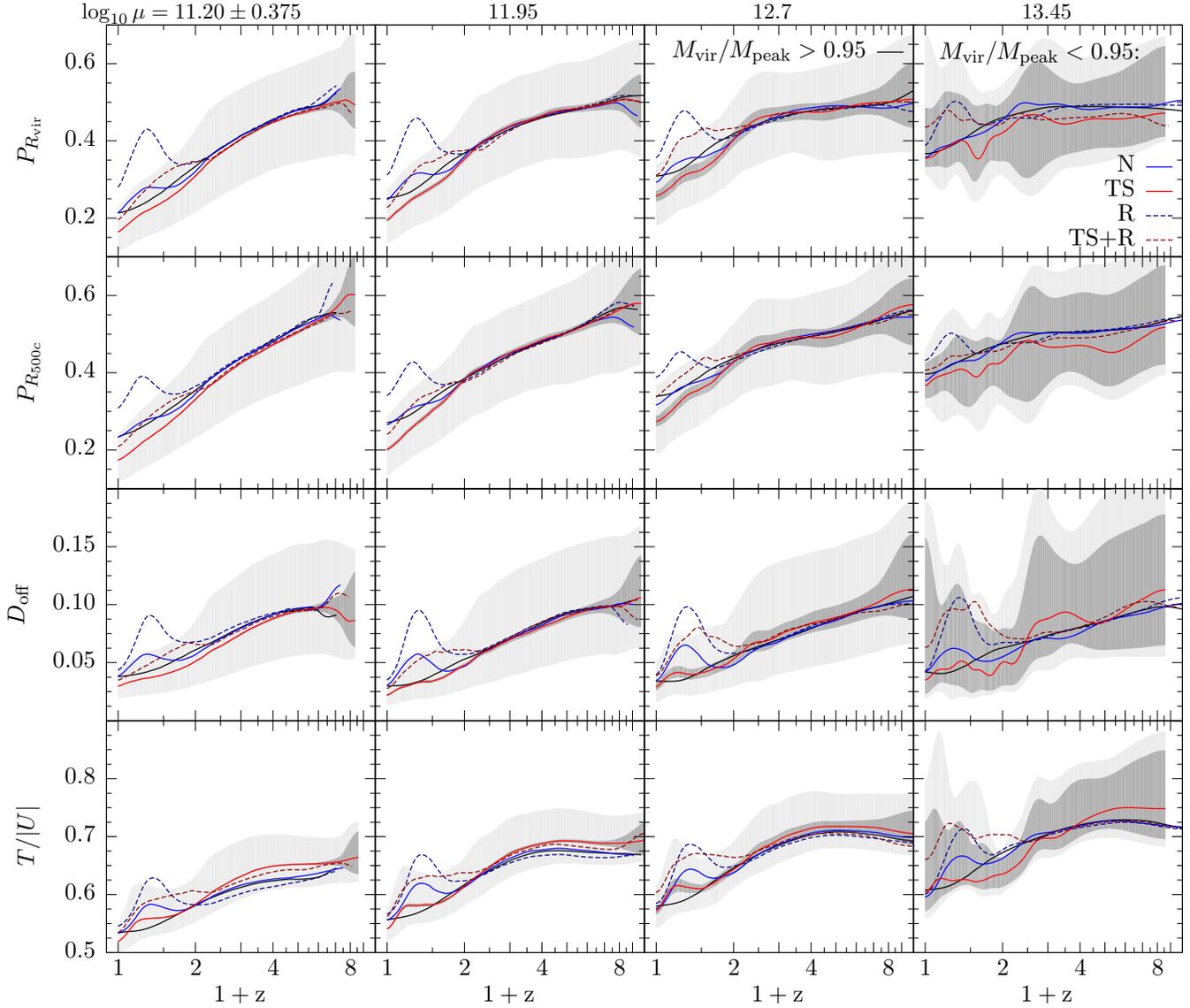}
    \caption{Same as Fig. \ref{fig:stripping_prog_hist_1}, but showing prolateness measured at $\rvir$ ($\pvir$), prolateness measured at $R_{500c}$ ($P_{R_{500c}}$), $\doff$, and the virial ratio. We see that halos experiencing purely tidal mass loss (Group TS) typically become steadily rounder and more symmetric.  In contrast, halos that underwent purely major merger induced mass loss (Group R) temporarily become highly elongated, especially at larger radii, highly asymmetric, and unrelaxed, all coincident with the merging event.  Group TS+R halos display much milder trends consistent with both tidal stripping and relaxation, while Group N halos display trends that are most consistent with those from Group R, but subdued, suggesting these may be halos experiencing weaker mass loss from minor mergers.  Diminished halos from all groups experienced a jump or plateau in virial ratio around $z = 0.5$, followed by relaxation by $z = 0$, with the most unrelaxed halos coming from groups R, TS+R, N, and TS, in order of most to least, respectively.}
    \label{fig:stripping_prog_hist_3}
\end{figure*}

\begin{figure*}
	\centering
	\includegraphics[trim=57 167 112 158, clip, width=\textwidth]{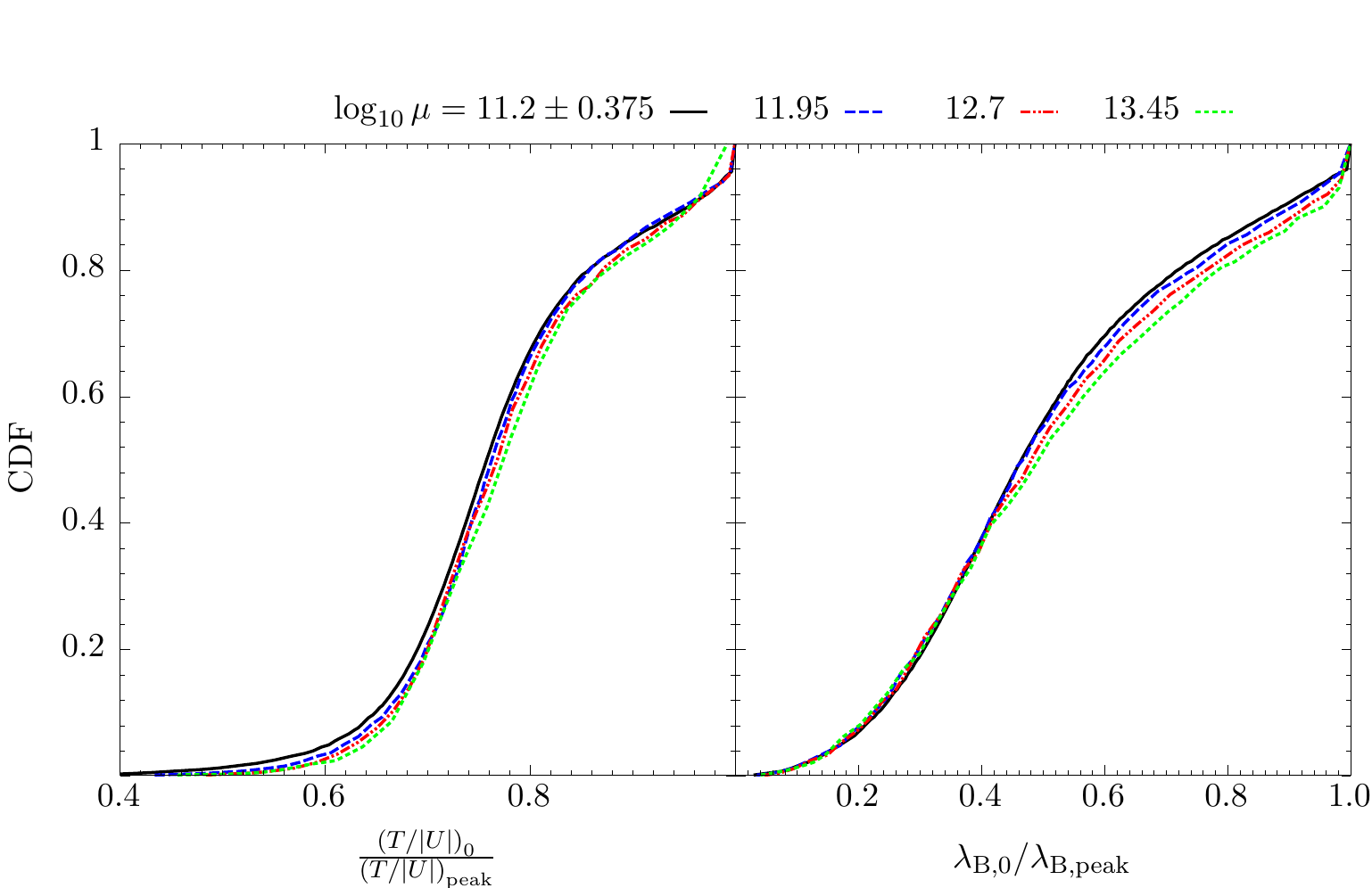}
    \caption{Cumulative distribution functions of virial ratio ($\tu$) decline on the left panel and spin parameter ($\lambdap$) decline on the right panel for all group R halos (distinct halos that had a recent major merger but have not experienced strong tidal forces). {\color{black}We use the same mass bin definitions as in Fig. \ref{fig:stripping_cdf}.} In each case, we compute the ratio of the $z = 0$ values to the peak value since the last major merger.  This provides an indication of the degree to which virial ratio and spin parameter typically decay following a major merger, since both of these properties tend to increase sharply immediately after a major merger.  Different coloured lines represent different mass bins, though we see very little mass dependence.  The median decline in virial ratio is about $24\%$ of the peak value, while the median decline in spin parameter is about $52\%$ of the peak value.  About $80\%$ of these group R halos decline by at least $15\%$ in virial ratio, while nearly $90\%$ decline by at least $15\%$ in spin parameter.  This tells us that elevated spin parameters following major mergers are transient; they typically decay substantially by $z = 0$, at least when considering only the material within the virial radius of the halo.  Fluctuations in virial ratio are less dramatic that in spin parameter, but show qualitatively similar time-dependence following a major merger.}
    \label{fig:spin_tu_cdf}
\end{figure*}

\begin{figure}
	\centering
	\includegraphics[trim=120 171 174 200, clip, width=\columnwidth]{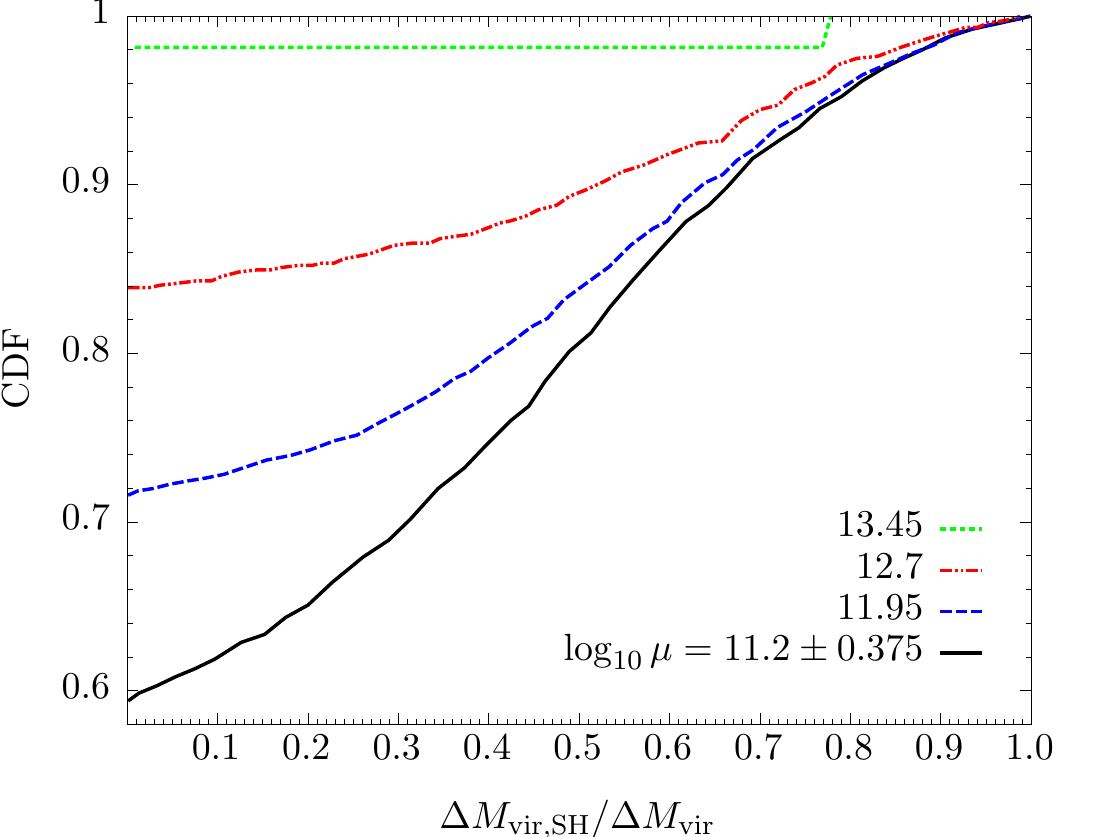}
    \caption{Cumulative distribution function of fraction of mass lost while a subhalo for all tidally stripped halos (group TS). {\color{black}We use the same mass bin definitions as in Fig. \ref{fig:stripping_cdf}.} We define $\Delta \mvir$ as the total cumulative mass loss since $\mpeak$, not including any periods of mass growth.  So $\Delta M_{\mathrm{vir,SH}}$ represents any loss that occurs while (temporarily) a subhalo.  Different coloured lines represent different mass bins.  For the lowest mass bin, we see that about $60\%$ of halos do not lose any of their mass as subhalos.  About $80\%$ of these halos lost half of their mass or less as a subhalos, leaving only about $20\%$ of halos having experienced the majority of their mass loss as a subhalo.  Mass loss while a subhalo is much less common for higher halo masses.  Roughly $72\%$ of $\log_{10} \mu = 11.95$ halos, $84\%$ of $\log_{10} \mu = 12.7$, and $98\%$ of $\log_{10} \mu = 13.45$ halos have not lost any mass as a subhalo.}
    \label{fig:mass_lost_as_sh_cdf}
\end{figure}

From Fig. \ref{fig:pdf_mechanisms2} we see that low mass diminished group TS and TS+R halos have a dramatically different distribution of mass loss ratio compared to group R and N halos.  While halos from all groups are more likely to be less diminished than more diminished, group TS and TS+R halos are much more likely to be heavily diminished ($\mvir/\mpeak < 0.7$) than group R and N halos.  Since the distributions of TS and TS+R halos are nearly identical (with only a difference of overall abundance), we expect that tidal stripping is the dominant mass loss mechanism affecting group TS+R halos, rather than post-merger mass loss.  The distributions of $\xoff$ are not remarkable for group TS and TS+R halos, with only a slight tendency towards lower $\xoff$ for group TS halos (more symmetric).  Group TS+R halos are about as relaxed as non-diminished halos, while the purely tidally stripped group TS halos tend to be more relaxed (at least, with lower $\tu$) than any of the other groups.  This isn't surprising, since tidal stripping should be a relatively distributed process, affecting mostly the outer regions of the halo and especially any loosely bound, high energy particles on highly elliptical orbits.  As has been repeated before, these trends hold especially for the lower mass halos, and begin to break down or lose statistical weight at higher masses.

Examining the evolution of halo properties for diminished group TS and TS+R halos further clarifies the role tidal stipping plays in shaping key halo properties like concentration, spin and shape, among others.  In Fig. \ref{fig:stripping_prog_hist_1} we see that low mass diminished group TS and TS+R halos develop increasingly higher concentrations, which diverge from those of the non-diminished halos around $z = 1$.  We expect this is roughly the epoch when these halos enter high density regions and begin to experience very high tidal forces.  Indeed, shortly following the divergence in concentration, the median mass of group TS and TS+R halos plateaus and begins to decline in response to tidal stripping.  Group TS halos have consistently higher concentrations than group TS+R halos since $z = 1$, while the median concentration of group TS+R halos initially drops below that of non-diminished halos, before rapidly increasing and surpasing them by $z = 0$.  These fluctuations may be due to the combination of early concentration suppression by a major merger, followed by concentration amplification from tidal stripping.  The spin parameters of diminished group TS halos follow a distinct trend compared to all other groups shown.  At least for the lower mass halos, those experiencing purely tidal mass loss also demonstrate increasingly reduced spin parameters.  While all halo populations had similar median spins above roughly $z = 2$, those in group TS diverge sharply from the non-diminished halos shortly after $z = 0.5$, around the same time the specific mass accretion rate reaches a minimum.  We interpret this to be a result of preferential removal of high energy material on highly elliptical orbits, reducing the net angular momentum of the halo.  Interestingly, group TS+R halos have a fairly different spin parameter evolution than group TS halos, even for the lowest mass bin, seen by the increase in spin parameter around $z = 1$.  This is presumably in response to a major merger, and results in a broader peak in spin parameter compared to that experienced by group R and N halos. For the specific mass accretion rate (bottom row), we see that for low mass group TS halos the net accumulation of material drops below zero shortly before $z = 0.5$ and reaches a minimum shortly before $z = 0$, but never substantially recovers.  Group TS+R halos display similar behaviour, but additionally experience the characteristic bump in accretion rate from a major merger typically around $z = 1$.  Note that the typical last major merger experienced by group TS+R halos occurs around $z = 1$, while for group R halos it peaks around $z = 0.5$.  This is likely due to the dominance of tidal stripping in the late phase of these halos' evolution.  Mergers are less likely to happen when a halo is being tidally disrupted by a larger halo, which places a constraint on the major merger window for group TS+R halos roughly between the assignment cutoff of $a = 0.45$ (or $z = 1.2$) and when stripping kicks in (around $z = 0.5$ for low mass halos).  It is clear from these plots that high mass group TS halos behave differently than low mass halos.  In fact, the evolution of high mass group TS halos (and especially group TS+R halos) approach the behaviour of group R halos; that is, a jump in spin parameter and reduced concentration coincident with an increase in accretion rate, followed by relaxation and negative accretion.  This suggests that high mass group TS and TS+R halos may be more strongly influenced by minor and major mergers, respectively, while low mass group TS and TS+R halos are likely dominated by the effects of tidal stripping.

By definition, group TS and TS+R halos have completely different tidal force histories, as seen in Fig. \ref{fig:stripping_prog_hist_3} Row 1.  However, note that the tidal forces experienced by halos in all groups are roughly comparable for $z > 2$, indicating that halos being tidally stripped in high density regions at $z = 0$ originated from regions of roughly average density where they experienced much milder tidal forces.  Around $z = 1$ the tidal forces experienced by diminished group TS and TS+R halos diverges sharply from the remaining groups until reaching a maximum around $\tf = 1.2$ near $z = 0$.  For lower mass halos in particular, both TS and TS+R halos peak shortly before $z = 0$, at around $z = 0.2$, roughly coincident with the minimum mass accretion rate (peak mass loss rate).  This tells us that specific mass accretion rate is strongly correlated with the tidal force, and that these halos typically have already endured a period of closest approach to a massive halo (peak in tidal force) paired with maximum instantaneous mass loss.  Presumably, most of these halos have already made one pass through or near a massive neighbour and are preparing to make subsequent passes before becoming subhalos.  Note that since we are only including ($z = 0$) distinct halos in our analysis, we are likely missing transient subhalos (halos that temporarily become subhalos for a period of time while passing through a massive halo, before becoming tidally stripped distinct halos upon emerging).  Some of the $z = 0$ distinct group TS and TS+R halos may have lost much of their mass as transient subhalos around the $z = 0.2$ peak in TF.  High mass group TS and TS+R halos still have very high tidal forces at late times, but do not show any clear indication of previous peaks in tidal force.

In Fig. \ref{fig:stripping_prog_hist_2} Rows 3 and 4 we show the evolution of the NFW scale radius and the maximum circular velocity ($\vmax$).  Low mass diminished group TS halos exhibit a dramatic reversal in scale radius evolution at about $z = 1$, roughly coincident with the transition towards higher density, higher tidal forces regions. Evolution in scale radius is the primary differentiator of evolution in concentration, since the virial radius is not highly sensitive to mass loss or group assignment.  For group TS halos, we see that the typical halo experiences a plateauing of scale radius around $z = 1$, followed by a sharp decline all the way to $z = 0$.  Group TS+R halos initially experience an increase in scale radius (presumably to due the major merger), peaking around $z = 0.5$, followed by a sharp decrease.  Tidal stripping preferentially removes loosely bound material from the outer regions of halos.  As a result, tidally stripped halos tend to have density profiles with outer slopes that fall off faster than $r^{-3}$.  Forcing an NFW fit to these halos produces artificially low scale radii in an attempt to compensate for the steep outer profiles.  {\color{black}Some of the internal halo structure is influenced by tidal stripping as well, as can be seen by the decline in $\vmax$ after $z = 1$.  $\vmax$ is not as sensitive to changes to the outer halo as NFW concentration, and provides a more robust quantifier of halo internal structure than the scale radius for non-NFW halos.  A decline in $\vmax$ indicates the removal of particles that spend some of their time in the interior of the halo.}  High mass group TS halos exhibit little to no scale radius and $\vmax$ suppression at late times, while high mass group TS+R halos have scale radius and $\vmax$ evolution comparable to that of group R halos, a further indication that high mass group TS+R halos are primarily influenced by the effects of a major merger rather than tidal stripping.

Finally, in Fig. \ref{fig:stripping_prog_hist_3}, we examine prolateness, asymmetry ($\doff$), and virial ratio ($\tu$).  All halos generally become less prolate with time, but group TS halos do so the most quickly on average.  While the median prolateness of halos in all groups is indistinguishable above $z = 1$, group TS halos sphericalize the most rapidly afterwards.  Still, the slope difference between group TS and group N halos after $z = 1$ is slight, and there does not appear to be a sharp change in $\pvir$ or $\pfive$ coincident with the onset of tidal stripping, as we see with spin and scale radius, for example.  Instead, tidally stripped halos are gradually rounded as they lose particles on highly elliptical orbits.  Both the inner ($\pfive$) and outer ($\pvir$) shapes of group TS halos are noticeably rounder than halos from other groups at $z = 0$, indicating that stripping is not solely affecting the outer regions of halos.  We also see that group TS halos are the most symmetric (lowest $\doff$) and have the lowest virial ratios at $z = 0$.  However, tidal stripping does not appear to strongly affect the halo center of mass, since there remains little difference between group TS and not-diminshed halos throughout their evolution.  Group TS halos do experience somewhat different virial ratio evolution, however, since they are typically the least relaxed at high redshift and experience a plateau during the tidal stripping phase.  Group TS+R halos have trends that combine elements of those from groups R and TS, favoring group TS and low masses and group R at high masses; that is, a jump up in prolateness (especially $\pvir$), $\doff$, and $\tu$, followed by gradual declines for low mass halos, and evolution consistent with group R halos at high masses.

Altogether, these trends produce a clear picture of how tidal stripping influences halo properties.  Group TS and TS+R halos move from average density regions at high redshift into high density regions where they experience increasingly strong tidal forces around $z = 1$.  These strong tidal forces are likely due to a single nearby massive halo, whose tidal field eventually begins to preferentially remove loosely bound high energy material from the subjected halo.  As the halo begins to loose mass due to tidal harassment, its scale radius begins to artificially depress due to a steepening outer profile, its spin decreases, and it becomes rounder.  Most halos reach a minimum accretion rate as they pass by or through the neighbouring halo shortly before $z = 0$.  We don't expect many of these group TS halos will remain distinct halos for long, since most are at their minimum mass since $\mpeak$ at $z = 0$ (i.e. very few have started accreting mass again after being stripped).  Group TS+R halos have a similar pattern of behaviour to group TS halos at late times, but additionally experienced a major merger (typically around $z = 1$, right before experiencing very strong tidal forces).  These halos initially evolve as group R halos do, with decreased concentrations and increased spins, accretion rates, prolateness, asymmetry, and virial ratios, before following the trends of group TS halos as tidal stripping commences.  We note that the trends outlined here apply most directly to low mass halos.  Higher mass halos seem to be increasingly influenced by merger events.  The evolution of high mass group TS+R halos appears dominated by the affects of recent major mergers.  Even high mass group TS halos may have lost mass via minor mergers rather than significant tidal stripping.

\section{Discussion} 

{\color{black}
By analysing the properties of group R halos that have lost more than $5\%$ of their peak mass at $z =0$ (as we did in Figs.~\ref{fig:pdf_mechanisms1} - \ref{fig:stripping_prog_hist_3}), we select 
halos whose last major merger 
occurred roughly at $a = 0.7$.  
Halos that had a major merger at a much earlier epoch likely already completed any potential mass loss phase and resumed accreting beyond their previous peak mass, while those that had a major merger much after $a = 0.7$ may still be accumulating material from the merger and have not yet begun to lose mass.  As a result, we don't follow the evolution of these diminished group R halos to the point that they typically start accreting normally again.  We get a glimpse of how the properties of an example individual halo recover after a major merger in Fig. \ref{fig:PD_R}; in this case, the effects of the merger are largely transient, with scale radius, shape, spin parameter, $\xoff$ and virial ratio all eventually returning to typical pre-merger values.  Spin parameter is one of the slowest to settle, and $\vmax$ never settles substantially.  In order to get a sense for how much the 
spin parameter and virial ratio typically decline after being elevated via a major merger, we examine in Fig. \ref{fig:spin_tu_cdf} the ratio of the $z = 0$ value of these quantities to their maximum values after $\mpeak$ for all group R halos.  Note we don't make any cuts on mass loss, though we expect most of these halos have lost mass at some point (see, e.g., Fig. \ref{fig:bsr_group_cdf}).  We see that the virial ratio of 
group R halos typically declines by about $25\%$ from their peak virial ratio after a major merger.  For a fully relaxed final halo, this would be a peak virial ratio of $\tu \sim 0.66$.  Very few halos decline by more than $40\%$, while about $80\%$ of halos decline by at least $15\%$.  Declines in spin parameter are more substantial, with the typical $z = 0$ value being about $52\%$ less than the peak value since the major merger.  The distribution is also more broad, with roughly $8\%$ of halos declining by over $80\%$ and only $10\%$ losing $15\%$ or less of their peak spin parameter value. Remarkably, there is very little mass dependence, indicating 
that 
major mergers affect halos in a relatively mass-independent manner.  Some of these halos will still be in the process of settling in virial ratio and spin parameter, tending to bias these results towards lower settling fractions.


While we focus on $z = 0$ distinct halos for our analysis, it is true that many tidally stripped halos have previously been subhalos for one (or more) short periods.  In the top row of Fig. \ref{fig:bsr_group_cdf} we include a curve for the lowest mass bin only that represents the cumulative distribution of mass loss fraction for halos that have never been subhalos (this dashed black line, normalized using the full mass bin).  From this we see that the majority of diminished group TS halos were subhalos previously, presumably on their first pass through a more massive halo (as was the case in the example in 
Fig. \ref{fig:PD_TS}).  Only about $35\%$ of diminished group TS halos have never been subhalos, and this fraction drops precipitously for halos that have lost more than about $20\%$ of their peak mass.  This tells us that tidally stripped distinct halos typically have already passed through a more massive halo, and are destined to be permanently captured as subhalos on the next approach or experience multiple additional pass-throughs before finally being captured.  The group TS halos that have lost mass but have never been subhalos likely experienced encounters where they passed nearby, but not within the virial radius of a larger halo.  The tidal influence of a massive halo can extend well beyond its virial radius; in fact, the tidal force quantifier we're using ($R_{\mathrm{Hill}}/\rvir$) typically rises above unity before a halo enters within the virial radius of a larger halo.  One further question we've investigated is whether the majority of tidal mass loss occurs while halos are passing through larger halos or while they are distinct halos (before or after passing through).  In Fig. \ref{fig:mass_lost_as_sh_cdf}, we show the distribution of the fraction of mass lost while a subhalo since $\mpeak$ for each of the four halo mass bins (group TS only).  Here we do not segregate halos based on mass loss severity, so this includes group TS halos that have lost very little to zero mass.  We compute $\Delta\mvir$ as the integrated mass loss, not including any positive contributions from periods of accretion, rather than the net mass loss.  We see a strong mass dependence; about $60\%$ of the lowest mass halos have not lost any mass as a subhalo, while virtually no high mass halos have lost mass as subhalos.  Roughly $20\%$ of halos in our lowest mass bin have lost half or more of their mass as subhalos.  Given that a majority of low mass diminished group TS halos have previously been subhalos, this implies that most of these halos experience the majority of their mass loss after they have passed through a more massive halo.  This picture is also consistent with our individual halo example (Fig. \ref{fig:PD_TS}), where most of the mass loss occurs after the halo has re-emerged from the larger halo (after the end of the thick purple line segment on the property evolution plots).


Throughout this work we've suggested that diminished group N halos may be the result of mass loss following minor (rather than major) mergers.  Indeed, the trends shown in Figs. \ref{fig:pdf_mechanisms1} - \ref{fig:stripping_prog_hist_3} illustrate convincing parallels between diminished group R and N halos.  A related analysis by students working with us 
{\color{black} \citep[][]{WuZhang17}} investigates halo by halo whether diminished group N halos did have recent minor mergers, as well as further characterize the responses of several halo properties to major (and minor) mergers.  They use the evolution of halo mass and $\vmax$ to predict the occurrence of mergers since $a = 0.5$ for each halo, validating their results with the known major merger events from the \rockstar\ catalog.  While this method may not be a reliable predictor of true minor merger events for an individual halo, it remains useful to provide statistics for populations as a whole.  In particular, they found that group N halos typically had their last minor merger around $a = 0.7$, consistent with the distribution of $\almm$ for group R halos, solidifying our conjecture that group N halos are an extension of the group R mass loss phenomenon towards smaller mass ratios.  They also build on the characterization of how spin parameter, $\xoff$, scale radius, prolateness, and virial ratio respond to major mergers (e.g, Figs. \ref{fig:stripping_prog_hist_2}-\ref{fig:stripping_prog_hist_3}), by for each property providing statistics on the number of peaks and when they occur following mergers.  Consistent with the results presented in this work, they show that spin parameter and virial ratio typically peak once about $\Delta a = 0.03$ after a merger, while $\xoff$, scale radius, and prolateness often have two merger-induced peaks (a result of the merging and backsplash of two separate high density halo cores).  For properties that peak twice, the first peak usually occurs immediately following the merger, while the second peak is typically delayed by about $\Delta a = 0.08$, but has a fairly broad distribution.  Furthermore, they find that the presence of two peaks in these properties is most common for halos that experience a maximum mass loss of $5-15\%$ of their peak mass (as opposed to only $<5\%$ or $>15\%$).  They also considered merger-induced 3rd and 4th peaks, but did not find any convincing indication that these occur.  All of these trends are roughly consistent for both major and minor mergers.


\section{Conclusions}  

Our main conclusions are as follows:

\begin{enumerate}

\item Roughly $22\%$ of low mass ($12\%$ of high mass) $z = 0$ distinct halos have lost more than $5\%$ of their peak virial mass.
\item Mass loss occurs either via tidal stripping in high density regions or via relaxation following a merger.
\item Relaxation after most major mergers results in 
 more than $5\%$ mass loss, with the regime of $5-15\%$ mass loss being the most common.  This is roughly true for all halo masses.
\item Merger-induced mass loss that peaks at $z = 0$ ($a = 1$) results from mergers around $z = 0.4$ ($a = 0.7$).  This is the characteristic delay between a merger and the minimum mass the halo subsequently reaches.  Note that this delay is somewhat time dependent and will be different for mergers that occur at different times.
\item Halos undergoing merger-induced mass loss typically have lower concentrations, higher spin parameters, are more elongated, more asymmetric, and less relaxed than halos not currently experiencing mass loss.  These differences are the result of strong impulses generated by a merger event that have not fully settled back to typical pre-merger values.
\item Minor mergers can also induce mass loss.  Minor merger induced mass loss parallels major merger induced mass loss, but has a generally weaker effect on halo properties.
\item A majority of low mass halos in high tidal force regions have lost more than $15\%$ of their peak mass and will not recover.  Significant amounts of mass loss ($>30\%$) are not uncommon.  High mass halos rarely experience tidal stripping.
\item Halos undergoing tidal stripping typically have higher concentrations, lower spin parameters, and are more spherical than halos not currently experiencing mass loss.  These differences result from steepening of the outer density profiles of halos via preferential removal of high energy material on elliptical orbits.
\item Most tidally stripped distinct halos that have lost more than $5\%$ of their peak mass were previously subhalos as they passed through a more massive halo.  This is more likely to have been the case for halos that have lost more mass.

\end{enumerate}}

\section*{Acknowledgements} We acknowledge stimulating discussions with Vladimir Avila-Reese, Sandra Faber, David Koo, and Frank van den Bosch.
CTL and JRP were supported by CANDELS grant HST GO-12060.12-A, provided by NASA through a grant from the Space Telescope Science Institute (STScI), which is operated by the Association of Universities for Research in Astronomy,  Incorporated, under NASA contract NAS5-26555.  PB was partially supported by a Giacconi Fellowship from STScI.  The remainder of support for PB through program number HST-HF2-51353.001-A was provided by NASA through a Hubble Fellowship grant from STScI. ARP was supported by UC-MEXUS Fellowship. AD was supported by NSF grants AST-1010033 and AST-1405962. 
AT and JZ participated in this research as high school students in the Scientific Internship Program at UCSC in summers 2015 and 2016, organized by Prof. Raja GuhaThakurta.
Computational resources supporting this work were provided by the NASA High-End Computing (HEC) Program through the NASA Advanced Supercomputing (NAS) Division at Ames Research Center, and by the Hyades astrocomputer system at UCSC.  


\bibliographystyle{mn2efix.bst}
\bibliography{newrefs}

\begin{thebibliography}{26}
\expandafter\ifx\csname natexlab\endcsname\relax\def\natexlab#1{#1}\fi

\bibitem[{{Behroozi} {et~al}\mbox{.}(2015){Behroozi}, {Knebe}, {Pearce},
  {Elahi}, {Han}, {Lux}, {Mao}, {Muldrew}, {Potter}, \& {Srisawat}}]{Notts}
{Behroozi} P. {et~al.}, 2015, \mnras, 454, 3020

\bibitem[{{Behroozi}, {Loeb} \& {Wechsler}(2013){Behroozi}, {Loeb}, \&
  {Wechsler}}]{BehrooziLoebWechsler}
{Behroozi} P.~S., {Loeb} A., {Wechsler} R.~H., 2013, \jcap, 6, 019

\bibitem[{{Behroozi} {et~al}\mbox{.}(2014){Behroozi}, {Wechsler}, {Lu}, {Hahn},
  {Busha}, {Klypin}, \& {Primack}}]{Behroozi14}
{Behroozi} P.~S., {Wechsler} R.~H., {Lu} Y., {Hahn} O., {Busha} M.~T., {Klypin}
  A., {Primack} J.~R., 2014, \apj, 787, 156

\bibitem[{{Behroozi}, {Wechsler} \& {Wu}(2013){Behroozi}, {Wechsler}, \&
  {Wu}}]{ROCKSTAR}
{Behroozi} P.~S., {Wechsler} R.~H., {Wu} H.-Y., 2013, \apj, 762, 109

\bibitem[{{Behroozi} {et~al}\mbox{.}(2013){Behroozi}, {Wechsler}, {Wu},
  {Busha}, {Klypin}, \& {Primack}}]{ConsTrees}
{Behroozi} P.~S., {Wechsler} R.~H., {Wu} H.-Y., {Busha} M.~T., {Klypin} A.~A.,
  {Primack} J.~R., 2013, \apj, 763, 18

\bibitem[{{Bryan} \& {Norman}(1998)}]{BryanNorman}
{Bryan} G.~L., {Norman} M.~L., 1998, \apj, 495, 80

\bibitem[{{Bullock} {et~al}\mbox{.}(2001){Bullock}, {Dekel}, {Kolatt},
  {Kravtsov}, {Klypin}, {Porciani}, \& {Primack}}]{Bullock+2001}
{Bullock} J.~S., {Dekel} A., {Kolatt} T.~S., {Kravtsov} A.~V., {Klypin} A.~A.,
  {Porciani} C., {Primack} J.~R., 2001, \apj, 555, 240

\bibitem[{{Dekel} {et~al}\mbox{.}(2013){Dekel}, {Zolotov}, {Tweed}, {Cacciato},
  {Ceverino}, \& {Primack}}]{Dekel13}
{Dekel} A., {Zolotov} A., {Tweed} D., {Cacciato} M., {Ceverino} D., {Primack}
  J.~R., 2013, \mnras, 435, 999

\bibitem[{{D'Onghia} \& {Navarro}(2007)}]{D'OnghiaNavarro07}
{D'Onghia} E., {Navarro} J.~F., 2007, \mnras, 380, L58

\bibitem[{{Frenk} \& {White}(2012)}]{FrenkWhite12}
{Frenk} C.~S., {White} S.~D.~M., 2012, Annalen der Physik, 524, 507

\bibitem[{{Hahn} {et~al}\mbox{.}(2009){Hahn}, {Porciani}, {Dekel}, \&
  {Carollo}}]{Hahn09}
{Hahn} O., {Porciani} C., {Dekel} A., {Carollo} C.~M., 2009, \mnras, 398, 1742

\bibitem[{{Hearin}, {Behroozi} \& {van den Bosch}(2016){Hearin}, {Behroozi}, \&
  {van den Bosch}}]{Hearin16}
{Hearin} A.~P., {Behroozi} P.~S., {van den Bosch} F.~C., 2016, \mnras, 461,
  2135

\bibitem[{{Klypin} {et~al}\mbox{.}(2016){Klypin}, {Yepes}, {Gottl{\"o}ber},
  {Prada}, \& {He{\ss}}}]{Klypin16}
{Klypin} A., {Yepes} G., {Gottl{\"o}ber} S., {Prada} F., {He{\ss}} S., 2016,
  \mnras, 457, 4340

\bibitem[{{Lee} {et~al}\mbox{.}(2017){Lee}, {Primack}, {Behroozi},
  {Rodr{\'{\i}}guez-Puebla}, {Hellinger}, \& {Dekel}}]{Lee17}
{Lee} C.~T., {Primack} J.~R., {Behroozi} P., {Rodr{\'{\i}}guez-Puebla} A.,
  {Hellinger} D., {Dekel} A., 2017, \mnras, 466, 3834

\bibitem[{{More}, {Diemer} \& {Kravtsov}(2015){More}, {Diemer}, \&
  {Kravtsov}}]{Splashback15}
{More} S., {Diemer} B., {Kravtsov} A.~V., 2015, \apj, 810, 36

\bibitem[{{Navarro}, {Frenk} \& {White}(1996){Navarro}, {Frenk}, \&
  {White}}]{NFW96}
{Navarro} J.~F., {Frenk} C.~S., {White} S.~D.~M., 1996, \apj, 462, 563

\bibitem[{{Peebles}(1969)}]{Peebles69}
{Peebles} P.~J.~E., 1969, \apj, 155, 393

\bibitem[{{Planck Collaboration} {et~al}\mbox{.}(2014){Planck Collaboration},
  {Ade}, {Aghanim}, {Armitage-Caplan}, {Arnaud}, {Ashdown}, {Atrio-Barandela},
  {Aumont}, {Baccigalupi}, {Banday}, \& et~al.}]{Planck13}
{Planck Collaboration} {et~al.}, 2014, \aap, 571, A16

\bibitem[{{Planck Collaboration} {et~al}\mbox{.}(2016){Planck Collaboration},
  {Ade}, {Aghanim}, {Arnaud}, {Ashdown}, {Aumont}, {Baccigalupi}, {Banday},
  {Barreiro}, {Bartlett}, \& et~al.}]{Planck15}
{Planck Collaboration} {et~al.}, 2016, \aap, 594, A13

\bibitem[{{Primack}(2012)}]{Primack12}
{Primack} J.~R., 2012, Annalen der Physik, 524, 535

\bibitem[{{Rodr{\'{\i}}guez-Puebla}
  {et~al}\mbox{.}(2016){Rodr{\'{\i}}guez-Puebla}, {Behroozi}, {Primack},
  {Klypin}, {Lee}, \& {Hellinger}}]{HaloDemographics}
{Rodr{\'{\i}}guez-Puebla} A., {Behroozi} P., {Primack} J., {Klypin} A., {Lee}
  C., {Hellinger} D., 2016, \mnras, 462, 893

\bibitem[{{Rodr{\'{\i}}guez-Puebla}
  {et~al}\mbox{.}(2017){Rodr{\'{\i}}guez-Puebla}, {Primack}, {Avila-Reese}, \&
  {Faber}}]{RP17}
{Rodr{\'{\i}}guez-Puebla} A., {Primack} J.~R., {Avila-Reese} V., {Faber} S.~M.,
  2017, \mnras, 470, 651

\bibitem[{{Somerville} {et~al}\mbox{.}(2017){Somerville}, {Behroozi}, {Pandya},
  {Dekel}, {Faber}, {Fontana}, {Huang}, {Koekemoer}, {Koo},
  {P{\'e}rez-Gonz{\'a}lez}, {Primack}, {Santini}, {Taylor}, \& {van der
  Wel}}]{Somerville17}
{Somerville} R.~S. {et~al.}, 2017, ArXiv e-prints

\bibitem[{{van den Bosch}(2017)}]{vandenBosch17}
{van den Bosch} F.~C., 2017, \mnras, 468, 885

\bibitem[{{Wechsler} {et~al}\mbox{.}(2002){Wechsler}, {Bullock}, {Primack},
  {Kravtsov}, \& {Dekel}}]{Wechsler02}
{Wechsler} R.~H., {Bullock} J.~S., {Primack} J.~R., {Kravtsov} A.~V., {Dekel}
  A., 2002, \apj, 568, 52

\bibitem[{{Wu} \& {Zhang}(2017)}]{WuZhang17}
{Wu} P., {Zhang} S., 2017, ArXiv e-prints

\end{thebibliography}








\bsp	
\label{lastpage}
\end{document}